%% file: main.tex
\documentclass[9pt]{extarticle}

\usepackage{microtype}
\usepackage[utf8]{inputenc}
\usepackage[T1]{fontenc}
\usepackage{eurosym}
\usepackage{url}
\usepackage{amssymb}
\usepackage{amsmath,amsfonts}
\usepackage{pifont}
\usepackage[square,numbers]{natbib}
\usepackage{colortbl}
\usepackage{multirow}
\usepackage{amsthm}
\usepackage{booktabs}
\usepackage{textcomp}
\usepackage[table,xcdraw]{xcolor}
\usepackage{fancyhdr}

\definecolor{cai_primary}{HTML}{4C9A99}
\definecolor{cai_secondary}{HTML}{307FE2}
\definecolor{cai_accent}{HTML}{1D8348}
\definecolor{cai_dark}{HTML}{3F4444}
\definecolor{human_color}{HTML}{173C47}

\definecolor{graph_lightcyan}{HTML}{B8D8D8}
\definecolor{graph_gray}{HTML}{E8F0EF}
\definecolor{graph_navy}{HTML}{2D5A56}
\definecolor{graph_arrow}{HTML}{3D7A79}
\definecolor{graph_accent}{HTML}{6BBFB5}
\definecolor{cai_light}{HTML}{F5F5F5}

\definecolor{bends_color}{HTML}{2C8C86}      
\definecolor{bends_light}{HTML}{7FC9C1}
\definecolor{mid_color}{HTML}{D98A2B}         
\definecolor{breaks_color}{HTML}{C0392B}      
\definecolor{cra_color}{HTML}{1D566E}         
\definecolor{cap_color}{HTML}{C0392B}         
\definecolor{alias_teal}{HTML}{4C9A99}        
\definecolor{agent_color}{HTML}{8E44AD}       

\newcommand{\bends}{\textcolor{bends_color}{\textbf{bends}}}
\newcommand{\breaks}{\textcolor{breaks_color}{\textbf{breaks}}}
\newcommand{\cra}{{\normalfont\scshape cra}}
\newcommand{\cai}{{\normalfont\scshape cai}}
\newcommand{\scw}[1]{{\normalfont\scshape #1}}   

\usepackage{graphicx}
\usepackage{subcaption}
\usepackage{float}
\usepackage{stfloats}
\usepackage{hyperref}
\hypersetup{
    colorlinks=true,
    urlcolor=cai_secondary,
    linkcolor=cai_secondary,
    filecolor=cai_accent,
    citecolor=cai_secondary,
}
\makeatletter
\g@addto@macro{\UrlBreaks}{\do\/\do\-\do\_\do\.\do\=\do\?\do\&}
\makeatother

\usepackage{tikz}
\newcommand{\premisebadge}[2]{
  \tikz[baseline=(p.base)]{\node[circle, fill=#1, text=white,
    font=\scriptsize\bfseries\sffamily, inner sep=0.9pt, minimum size=1.15em]
    (p) {#2};}}
\newcommand{\premise}[1]{\premisebadge{bends_color}{P#1}}      
\newcommand{\premiseb}[1]{\premisebadge{breaks_color}{P#1}}    

\usetikzlibrary{arrows.meta,positioning,shapes.geometric,calc,patterns,shadows,decorations.pathreplacing,fit,backgrounds}
\input{tex/robot_icons.tex}   
\usepackage{pgfplots}
\pgfplotsset{compat=1.16}
\usepgfplotslibrary{groupplots}

\pgfplotsset{
  every axis/.append style={
    label style={font=\small\sffamily, color=cai_dark},
    tick label style={font=\scriptsize\sffamily, color=cai_dark!88},
    title style={font=\footnotesize\sffamily, color=cai_dark, align=center},
    axis line style={draw=cai_dark!55, line width=0.6pt},
    every tick/.append style={draw=cai_dark!45, line width=0.5pt},
    tick align=outside,
    legend cell align=left,
    legend style={font=\scriptsize\sffamily, draw=cai_dark!22, fill=white,
      fill opacity=0.92, text opacity=1, rounded corners=2pt, inner sep=3pt, row sep=1pt},
    grid style={line width=0.4pt, draw=cai_dark!12},
  },
  every axis plot/.append style={mark options={fill opacity=0.95}},
}

\usepackage{enumitem}
\usepackage{ifthen}
\usepackage[breakable,skins]{tcolorbox}

\usepackage{geometry}
\renewcommand{\headrulewidth}{0.4pt}
\renewcommand{\footrulewidth}{0.4pt}
\renewcommand{\headrule}{\hbox to\headwidth{\color{cai_primary}\leaders\hrule height \headrulewidth\hfill}}
\renewcommand{\footrule}{\hbox to\headwidth{\color{human_color}\leaders\hrule height \footrulewidth\hfill}}
\newtheorem{prediction}{Prediction}
\newtheorem{definition}{Definition}

\usepackage{caption,setspace}
\usepackage{titlesec}
\titleformat{\section}
  {\normalfont\Large\bfseries\color{cai_primary}}{\thesection}{1em}{}[\titlerule]
\titleformat{\subsection}
  {\normalfont\large\bfseries\color{human_color}}{\thesubsection}{1em}{}
\titleformat{\subsubsection}
  {\normalfont\normalsize\bfseries\color{cai_dark}}{\thesubsubsection}{1em}{}

\usepackage{authblk}

\makeatletter
\let\orig@maketitle\maketitle
\renewcommand{\maketitle}{%
  \orig@maketitle%
  \vspace{-1.5em}%
  {\color{cai_primary!30}\hrule height 0.5pt}%
  \vspace{1em}%
}
\makeatother

\makeatletter
\renewenvironment{abstract}{\small\noindent\ignorespaces}{\par}
\makeatother

\begin{document}

\title{\LARGE\textcolor{cai_primary}{\textbf{Certifying Ghosts: How Cybersecurity AI \\ Agents Break the EU Cyber Resilience Act}}}

\author{Víctor Mayoral-Vilches}
\affil{\small Alias Robotics}
\date{}

\twocolumn[
\maketitle

\begin{abstract}
\noindent The EU Cyber Resilience Act (\cra{}, Regulation (EU) 2024/2847) makes a smart bet. It does not demand that products be free of vulnerabilities, a promise no software can keep, but only that manufacturers \emph{run a process}: assess risk, handle flaws, ship updates. The bet pays off only while four things about the world stay true: \premise{1}~finding vulnerabilities is slow, skilled, human work; \premiseb{2}~a product's exploitable flaws are knowable the day it ships; \premiseb{3}~exploitation is rare enough to notice; and \premiseb{4}~fixes keep pace with discovery. \emph{Cybersecurity AI (\cai{}) agents}, AI put to work finding and exploiting flaws in \emph{other} products, falsify all four. The regime answers in two opposite ways. Against the sheer \emph{volume} of flaws that agents surface it \emph{bends}~(\premise{1}): built for scarce attention, it re-centres compliance on defensible, documented prioritisation, and holds. But agents also \emph{collapse} the speed and economics of the vulnerability lifecycle, and here it \emph{breaks}~(\premiseb{2}\,\premiseb{3}\,\premiseb{4}): a product that passed every check becomes exploitable without anyone touching it, so its market-entry test, its reporting trigger, and its one-and-done certificate vouch for a security that has quietly expired. The fault is in the landscape, not the product, so running the process more diligently cannot repair it. Anchoring every legal claim in the enacted text and every capability claim in dated results, we map each mechanism to the force that strains or snaps it, and find the cure and the disease cut from the same cloth: because defenders and attackers wield the same AI, the only conformity that survives is one that never stops running. We then carry the remedy from proposal to proof on two \cra{}-scope robots, a humanoid and a lawn mower, where an agentic defender holds a line their undefended selves cannot. On the evidence already in hand, the \cra{} reaches full force in December~2027 certifying products against a world that has already left the building. Static, human-paced security is finished; what replaces it must be continuous and agent-operated, and that is no longer a matter of taste.
\end{abstract}
\vspace{-0.1em}
\begin{center}
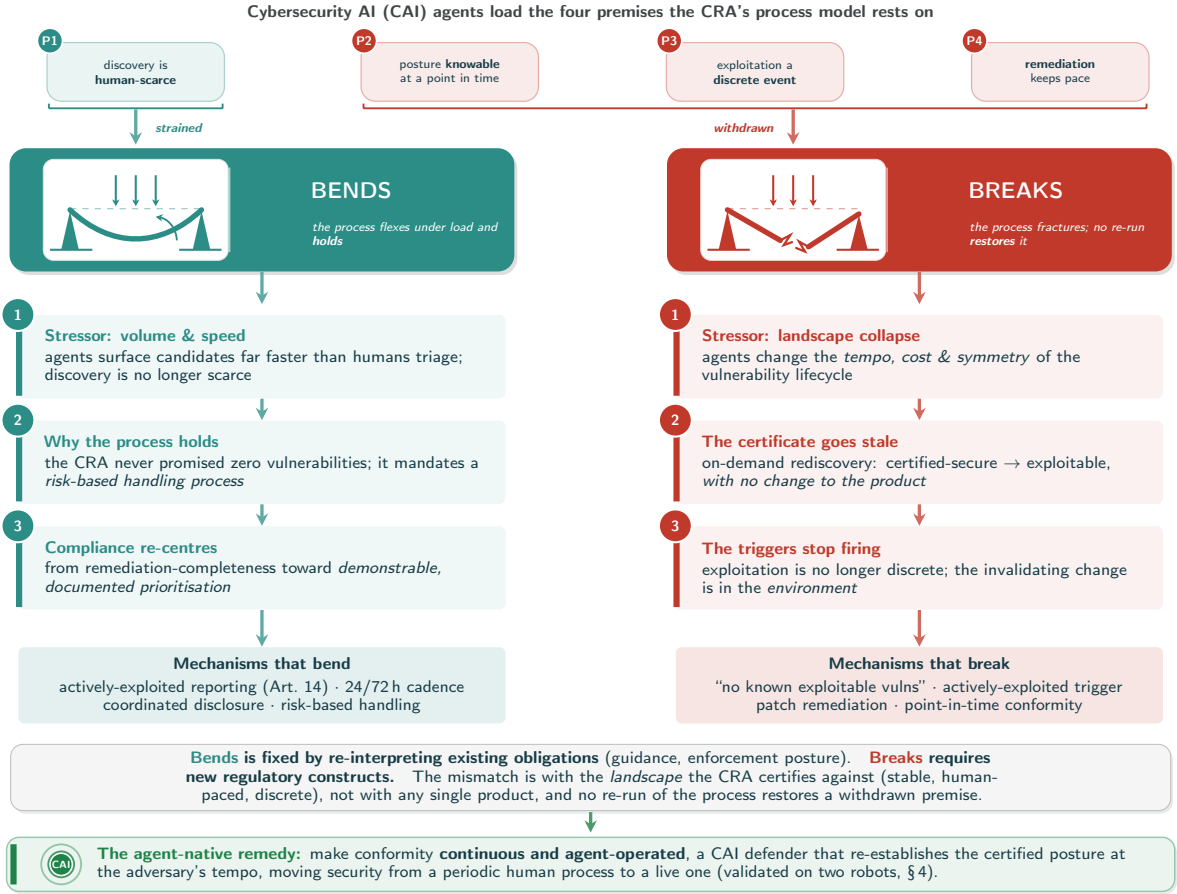

\resizebox{0.9\textwidth}{!}{\input{tex/fig_bends_breaks.tex}}
\end{center}
\vspace{0.0em}
\captionof{figure}{\textbf{The central distinction.} A process-oriented cybersecurity regime drawn as a loaded beam: under the \emph{same} rising load from Cybersecurity AI (\cai{}) agents it either \emph{bends} (flexes and holds; left, teal) or \emph{breaks} (fractures irreparably; right, red). Agents \emph{strain} premise~P1, so the regime bends and compliance re-centres on demonstrable prioritisation; they \emph{withdraw} P2--P4, so it breaks, because the mechanisms encoding those premises certify a posture that no longer persists. \scw{bends} problems yield to re-interpreting existing obligations; \scw{breaks} problems require new constructs, because the mismatch is with the landscape, not any single product. A final band names the resolution: an agent-native remedy that makes conformity \emph{continuous and agent-operated} (\S\,\ref{sec:solution}).}\label{fig:hero}
\vspace{0.6em}
]

\newpage

\section{Introduction}\label{sec:intro}
\input{tex/introduction.tex}

\section{The Process-Oriented Regime and Its Premises}\label{sec:background}
\input{tex/regime.tex}

\section{Disarming the CRA in the AI Era}\label{sec:core}
\input{tex/core.tex}

\section{An Agent-Native Remedy}\label{sec:solution}
\input{tex/remedy.tex}

\section{Conclusion}\label{sec:conclusion}
\input{tex/conclusion.tex}

\bibliography{bibliography}

\input{tex/appendix_zerodayclock.tex}

\input{tex/appendix_euvd.tex}

\input{tex/appendix_euroclock.tex}

\end{document}

%% file: tex/robot_icons.tex
\newcommand{\rcHumanoid}[3]{\begin{scope}[shift={(#1,#2)}, line cap=round, line join=round, draw=#3]
  \draw[line width=3.1pt, color=#3!55] (-0.09,-0.03) -- (-0.10,-0.26);
  \draw[line width=3.1pt, color=#3!55] (0.09,-0.03) -- (0.10,-0.26);
  \draw[line width=2.9pt, color=#3!55] (-0.10,-0.27) -- (-0.095,-0.47);
  \draw[line width=2.9pt, color=#3!55] (0.10,-0.27) -- (0.095,-0.47);
  \draw[fill=#3!68, draw=#3, line width=0.5pt, rounded corners=0.7pt] (-0.16,-0.505) rectangle (-0.028,-0.45);
  \draw[fill=#3!68, draw=#3, line width=0.5pt, rounded corners=0.7pt] (0.028,-0.505) rectangle (0.16,-0.45);
  \draw[line width=2.9pt, color=#3!55] (-0.17,0.37) -- (-0.235,0.16);
  \draw[line width=2.9pt, color=#3!55] (0.17,0.37) -- (0.235,0.16);
  \draw[line width=2.6pt, color=#3!55] (-0.235,0.15) -- (-0.225,-0.05);
  \draw[line width=2.6pt, color=#3!55] (0.235,0.15) -- (0.225,-0.05);
  \draw[fill=#3!14, draw=#3, line width=0.5pt] (-0.225,-0.085) circle (0.045);
  \draw[fill=#3!14, draw=#3, line width=0.5pt] (0.225,-0.085) circle (0.045);
  \draw[fill=#3!55, draw=#3, line width=0.6pt, rounded corners=1.3pt] (-0.135,-0.075) rectangle (0.135,0.055);
  \draw[fill=#3!13, draw=#3, line width=0.7pt, rounded corners=2.6pt] (-0.16,0.055) rectangle (0.16,0.405);
  \draw[draw=#3!42, line width=0.5pt, rounded corners=1pt] (-0.075,0.135) rectangle (0.075,0.30);
  \draw[#3!38, line width=0.4pt] (0,0.315) -- (0,0.405);
  \foreach \x/\y in {-0.17/0.375,0.17/0.375}{\draw[fill=#3!70,draw=#3,line width=0.4pt] (\x,\y) circle (0.053);}
  \foreach \x/\y in {-0.09/-0.03,0.09/-0.03}{\draw[fill=#3!70,draw=#3,line width=0.4pt] (\x,\y) circle (0.045);}
  \foreach \x/\y in {-0.10/-0.265,0.10/-0.265}{\draw[fill=#3!70,draw=#3,line width=0.4pt] (\x,\y) circle (0.04);}
  \draw[line width=2.6pt, color=#3!55] (0,0.40) -- (0,0.45);
  \draw[fill=#3!13, draw=#3, line width=0.7pt, rounded corners=3.2pt] (-0.12,0.45) rectangle (0.12,0.66);
  \draw[fill=#3!70, draw=#3!70, line width=0.5pt, rounded corners=1.6pt] (-0.085,0.515) rectangle (0.085,0.585);
\end{scope}}

\newcommand{\rcMower}[3]{\begin{scope}[shift={(#1,#2)}, line cap=round, line join=round, draw=#3]
  \draw[fill=#3!58, draw=#3, line width=0.6pt] (0.24,0.03) circle (0.13);
  \draw[fill=#3!16, draw=#3, line width=0.4pt] (0.24,0.03) circle (0.05);
  \draw[fill=#3!58, draw=#3, line width=0.6pt] (-0.26,0.01) circle (0.10);
  \draw[fill=#3!16, draw=#3, line width=0.4pt] (-0.26,0.01) circle (0.04);
  \draw[fill=#3!13, draw=#3, line width=0.7pt, rounded corners=2.2pt]
     (-0.40,0.05) -- (-0.40,0.13) -- (-0.24,0.235) -- (0.22,0.235) -- (0.40,0.14) -- (0.40,0.05) -- cycle;
  \draw[#3!45, line width=0.5pt] (-0.34,0.115) -- (0.34,0.125);
  \draw[fill=#3!70,draw=#3,line width=0.3pt] (-0.35,0.155) circle (0.022);
  \draw[#3!45, line width=0.5pt] (-0.14,0.0) -- (0.10,0.0);
  \draw[fill=#3!32, draw=#3, line width=0.5pt, rounded corners=0.6pt] (-0.05,0.235) rectangle (0.07,0.30);
  \draw[fill=#3!68, draw=#3, line width=0.5pt, rounded corners=1.6pt] (-0.10,0.30) rectangle (0.12,0.37);
  \draw[#3!22, line width=0.5pt] (-0.09,0.335) -- (0.11,0.335);
\end{scope}}

%% file: tex/fig_bends_breaks.tex
\begingroup\renewcommand{\cra}{CRA}\renewcommand{\cai}{CAI}%
\begin{tikzpicture}[
    every node/.style={font=\normalsize\sffamily, color=cai_dark},
    stage/.style 2 args={rounded corners=3pt,
        minimum width=8.3cm, text width=7.4cm, minimum height=1.42cm, align=left, fill=#1!7,
        text=human_color, inner xsep=12pt, inner ysep=7pt, font=\small\sffamily},
    accent/.style={line width=3pt, color=#1},
    badge/.style={circle, draw=white, line width=1pt, fill=#1, text=white,
        font=\footnotesize\bfseries\sffamily, minimum size=0.56cm, inner sep=0pt},
    mechchip/.style={rounded corners=3pt, fill=#1!13,
        minimum width=8.3cm, minimum height=1.3cm, align=center, text=human_color,
        font=\small\sffamily, inner xsep=8pt},
    arr/.style={-{Stealth[length=6pt,width=6pt]}, line width=1.6pt, color=#1!75},
    prempill/.style 2 args={rounded corners=7pt, fill=#1!10, draw=#1!40, line width=0.6pt,
        minimum width=2.95cm, text width=2.6cm, minimum height=1.0cm, align=center,
        text=human_color, font=\scriptsize\sffamily, inner xsep=6pt, inner ysep=3pt},
    pbadge/.style={circle, fill=#1, text=white, draw=white, line width=0.8pt,
        font=\scriptsize\bfseries\sffamily, inner sep=1.2pt, minimum size=0.42cm},
]
\def\LX{0}
\def\RX{11.2}
\def\bx{bends_color}
\def\kx{breaks_color}
\def\byc{1.95}   
\def\bxc{-2.15}  
\def\kxc{9.05}   

\node[font=\small\bfseries\sffamily, color=cai_dark, anchor=south] at (5.6,5.05)
  {Cybersecurity AI (\cai{}) agents load the four premises the \cra{}'s process model rests on};
\foreach \cx/\col/\pid/\gloss in {%
  -2.15/\bx/P1/{discovery is\\\textbf{human-scarce}},%
  3.2/\kx/P2/{posture \textbf{knowable}\\at a point in time},%
  8.4/\kx/P3/{exploitation a\\\textbf{discrete event}},%
  13.6/\kx/P4/{\textbf{remediation}\\keeps pace}%
}{
  \node[prempill={\col}] (\pid) at (\cx,4.30) {\gloss};
  \node[pbadge={\col}] at ($(\pid.north west)+(0.06,0.02)$) {\pid};
}

\draw[\bx, line width=0.9pt] (-3.63,3.68) -- (-0.67,3.68)
      (-3.63,3.68) -- (-3.63,3.76) (-0.67,3.68) -- (-0.67,3.76);
\draw[arr={\bx}] (\bxc,3.66) -- (\bxc,3.06);
\node[font=\scriptsize\itshape\bfseries\sffamily, text=\bx, anchor=west] at (\bxc+0.22,3.36) {strained};

\draw[\kx, line width=0.9pt] (1.73,3.68) -- (15.07,3.68)
      (1.73,3.68) -- (1.73,3.76) (15.07,3.68) -- (15.07,3.76);
\draw[arr={\kx}] (\kxc,3.66) -- (\kxc,3.06);
\node[font=\scriptsize\itshape\bfseries\sffamily, text=\kx, anchor=east] at (\kxc-0.22,3.36) {withdrawn};

\node[rounded corners=8pt, fill=\bx, minimum width=8.6cm, minimum height=2.10cm,
      drop shadow={shadow xshift=0.05cm, shadow yshift=-0.06cm, fill=cai_dark, opacity=0.18}]
  (bh) at (\LX,\byc) {};
\node[rounded corners=5pt, fill=white, draw=\bx!20, line width=0.5pt,
      minimum width=3.15cm, minimum height=1.72cm] at (\bxc,\byc) {};
\foreach \dx in {-0.34,0,0.34}{
  \draw[\bx, line width=1pt, -{Stealth[length=3.4pt,width=3.6pt]}]
    (\bxc+\dx,2.54) -- (\bxc+\dx,\byc+0.06);}
\draw[\bx!45, dashed, line width=0.6pt] (\bxc-1.12,\byc+0.02) -- (\bxc+1.12,\byc+0.02);
\draw[\bx, line width=2.6pt, line cap=round]
  (\bxc-1.12,\byc) .. controls (\bxc-0.42,\byc-0.66) and (\bxc+0.42,\byc-0.66) .. (\bxc+1.12,\byc);
\draw[\bx, line width=1pt] (\bxc-1.46,\byc-0.68) -- (\bxc-0.74,\byc-0.68);
\draw[\bx, line width=1pt] (\bxc+0.74,\byc-0.68) -- (\bxc+1.46,\byc-0.68);
\fill[\bx] (\bxc-1.12,\byc-0.02) -- (\bxc-1.27,\byc-0.68) -- (\bxc-0.97,\byc-0.68) -- cycle;
\fill[\bx] (\bxc+1.12,\byc-0.02) -- (\bxc+0.97,\byc-0.68) -- (\bxc+1.27,\byc-0.68) -- cycle;
\draw[\bx, line width=0.9pt, -{Stealth[length=3.4pt,width=3.6pt]}]
  (\bxc+0.70,\byc-0.52) to[bend right=40] (\bxc+0.34,\byc-0.12);
\draw[white, line width=0.7pt, opacity=0.6] (\LX-0.55,\byc-0.72) -- (\LX-0.55,\byc+0.72);
\node[anchor=west, text=white, font=\Large\bfseries\sffamily] at ($(bh.center)+(0.72,0.34)$) {BENDS};
\node[anchor=west, text=white, font=\scriptsize\itshape\sffamily, text width=3.4cm]
  at ($(bh.center)+(0.74,-0.40)$) {the process flexes under load and \textbf{holds}};

\node[rounded corners=8pt, fill=\kx, minimum width=8.6cm, minimum height=2.10cm,
      drop shadow={shadow xshift=0.05cm, shadow yshift=-0.06cm, fill=cai_dark, opacity=0.18}]
  (kh) at (\RX,\byc) {};
\node[rounded corners=5pt, fill=white, draw=\kx!20, line width=0.5pt,
      minimum width=3.15cm, minimum height=1.72cm] at (\kxc,\byc) {};
\foreach \dx in {-0.34,0,0.34}{
  \draw[\kx, line width=1pt, -{Stealth[length=3.4pt,width=3.6pt]}]
    (\kxc+\dx,2.54) -- (\kxc+\dx,\byc+0.06);}
\draw[\kx!45, dashed, line width=0.6pt] (\kxc-1.12,\byc+0.02) -- (\kxc+1.12,\byc+0.02);
\draw[\kx, line width=2.6pt, line cap=round] (\kxc-1.12,\byc) -- (\kxc-0.22,\byc-0.52);
\draw[\kx, line width=1.1pt, line cap=round]
  (\kxc-0.22,\byc-0.52) -- (\kxc-0.05,\byc-0.43) -- (\kxc-0.16,\byc-0.60) -- (\kxc-0.02,\byc-0.55);
\draw[\kx, line width=2.6pt, line cap=round] (\kxc+0.26,\byc-0.63) -- (\kxc+1.12,\byc-0.07);
\draw[\kx, line width=1.1pt, line cap=round]
  (\kxc+0.26,\byc-0.63) -- (\kxc+0.10,\byc-0.54) -- (\kxc+0.20,\byc-0.71) -- (\kxc+0.05,\byc-0.65);
\draw[\kx, line width=1pt] (\kxc-1.46,\byc-0.68) -- (\kxc-0.74,\byc-0.68);
\draw[\kx, line width=1pt] (\kxc+0.74,\byc-0.68) -- (\kxc+1.46,\byc-0.68);
\fill[\kx] (\kxc-1.12,\byc-0.02) -- (\kxc-1.27,\byc-0.68) -- (\kxc-0.97,\byc-0.68) -- cycle;
\fill[\kx] (\kxc+1.12,\byc-0.09) -- (\kxc+0.97,\byc-0.68) -- (\kxc+1.27,\byc-0.68) -- cycle;
\draw[white, line width=0.7pt, opacity=0.6] (\RX-0.55,\byc-0.72) -- (\RX-0.55,\byc+0.72);
\node[anchor=west, text=white, font=\Large\bfseries\sffamily] at ($(kh.center)+(0.72,0.34)$) {BREAKS};
\node[anchor=west, text=white, font=\scriptsize\itshape\sffamily, text width=3.4cm]
  at ($(kh.center)+(0.74,-0.40)$) {the process fractures; no re-run \textbf{restores} it};

\draw[arr={\bx}] (bh.south) -- ($(bh.south)+(0,-0.55)$);
\draw[arr={\kx}] (kh.south) -- ($(kh.south)+(0,-0.55)$);

\node[stage={\bx}] (b1) at (\LX,-0.55) {{\bfseries\color{\bx} Stressor: volume \& speed}\\[1pt] agents surface candidates far faster than humans triage; discovery is no longer scarce};
\node[stage={\bx}] (b2) at (\LX,-2.35) {{\bfseries\color{\bx} Why the process holds}\\[1pt] the \cra{} never promised zero vulnerabilities; it mandates a \emph{risk-based handling process}};
\node[stage={\bx}] (b3) at (\LX,-4.15) {{\bfseries\color{\bx} Compliance re-centres}\\[1pt] from remediation-completeness toward \emph{demonstrable, documented prioritisation}};
\node[mechchip={\bx}] (b4) at (\LX,-6.15) {\textbf{Mechanisms that bend}\\[2pt] actively-exploited reporting (Art.~14)~$\cdot$~24/72\,h cadence\\ coordinated disclosure~$\cdot$~risk-based handling};
\foreach \b in {b1,b2,b3}{\draw[accent={\bx}] (\b.south west)++(0.02,0.05) -- ($(\b.north west)+(0.02,-0.05)$);}
\foreach \a/\b in {b1/b2,b2/b3,b3/b4}{\draw[arr={\bx}] (\a) -- (\b);}
\foreach \b/\n in {b1/1,b2/2,b3/3}{\node[badge={\bx}] at (\b.north west) {\n};}

\node[stage={\kx}] (k1) at (\RX,-0.55) {{\bfseries\color{\kx} Stressor: landscape collapse}\\[1pt] agents change the \emph{tempo, cost \& symmetry} of the vulnerability lifecycle};
\node[stage={\kx}] (k2) at (\RX,-2.35) {{\bfseries\color{\kx} The certificate goes stale}\\[1pt] on-demand rediscovery: certified-secure $\to$ exploitable, \emph{with no change to the product}};
\node[stage={\kx}] (k3) at (\RX,-4.15) {{\bfseries\color{\kx} The triggers stop firing}\\[1pt] exploitation is no longer discrete; the invalidating change is in the \emph{environment}};
\node[mechchip={\kx}] (k4) at (\RX,-6.15) {\textbf{Mechanisms that break}\\[2pt] ``no known exploitable vulns''~$\cdot$~actively-exploited trigger\\ patch remediation~$\cdot$~point-in-time conformity};
\foreach \b in {k1,k2,k3}{\draw[accent={\kx}] (\b.south west)++(0.02,0.05) -- ($(\b.north west)+(0.02,-0.05)$);}
\foreach \a/\b in {k1/k2,k2/k3,k3/k4}{\draw[arr={\kx}] (\a) -- (\b);}
\foreach \b/\n in {k1/1,k2/2,k3/3}{\node[badge={\kx}] at (\b.north west) {\n};}

\node[draw=cai_dark!35, fill=cai_light, rounded corners=6pt, align=center,
      text width=19.2cm, minimum height=1.10cm, font=\small\sffamily, text=human_color, inner xsep=8pt,
      drop shadow={shadow xshift=0.04cm, shadow yshift=-0.05cm, fill=cai_dark, opacity=0.12}]
  at (5.6,-7.72)
  {\textbf{\textcolor{\bx}{Bends} is fixed by re-interpreting existing obligations} (guidance, enforcement posture).\quad
   \textbf{\textcolor{\kx}{Breaks} requires new regulatory constructs.}\quad
   The mismatch is with the \emph{landscape} the \cra{} certifies against (stable, human-paced, discrete), not with any single product, and no re-run of the process restores a withdrawn premise.};

\draw[-{Stealth[length=6pt,width=6pt]}, line width=1.6pt, color=cai_accent!80] (5.6,-8.31) -- (5.6,-8.66);
\node[rounded corners=6pt, fill=cai_accent!9, draw=cai_accent!45, line width=0.7pt,
      minimum width=19.9cm, minimum height=0.92cm,
      drop shadow={shadow xshift=0.04cm, shadow yshift=-0.05cm, fill=cai_dark, opacity=0.12}]
  (soln) at (5.6,-9.20) {};
\draw[cai_accent, line width=3pt] ($(soln.north west)+(0.13,-0.13)$) -- ($(soln.south west)+(0.13,0.13)$);
\foreach \r/\op in {0.34/0.42, 0.22/0.8}{\draw[cai_accent, line width=0.9pt, opacity=\op] (-3.42,-9.20) circle (\r);}
\node[circle, fill=cai_accent, text=white, draw=white, line width=0.5pt,
      font=\tiny\bfseries\sffamily, inner sep=0, minimum size=3.7mm] at (-3.42,-9.20) {\cai{}};
\node[anchor=west, align=left, text width=17.7cm, font=\small\sffamily, text=human_color] at (-2.92,-9.20)
  {\textbf{\textcolor{cai_accent}{The agent-native remedy:}} make conformity \textbf{continuous and agent-operated}, a \cai{} defender that re-establishes the certified posture at the adversary's tempo, moving security from a periodic human process to a live one (validated on two robots, \S\,4).};
\end{tikzpicture}\endgroup

%% file: tex/introduction.tex
The European Union's Cyber Resilience Act (\cra{}), Regulation (EU) 2024/2847, is the first horizontal law to impose mandatory cybersecurity requirements on products with digital elements placed on the Union market~\citep{cra2024}. It entered into force on 10~December 2024; its central vulnerability- and incident-reporting obligation (Article~14) applies from 11~September 2026, and the bulk of its substantive obligations from 11~December 2027 (Article~71). The design is deliberate, and, for the world it was drafted in, astute. Rather than demand that products be free of vulnerabilities, an outcome no non-trivial software can guarantee, the \cra{} requires that manufacturers \emph{run a process}: assess and document cybersecurity risk, handle vulnerabilities across a declared support period, ship security updates, exercise supply-chain due diligence, and report vulnerabilities under active exploitation. The regime is process-oriented, not outcome-oriented.

This paper argues that this design has already been overtaken by the vulnerability ecosystem that generative AI and Cybersecurity AI (\cai{}) agents are now producing, and that the failure is present, not hypothetical. The point is sharp because the \cra{}'s foundational assumptions were fixed at a specific moment: the Commission published the proposal on 15~September 2022~\citep{cra_proposal}, and OpenAI released ChatGPT on 30~November 2022. The \cra{}'s model of how vulnerabilities are discovered, predominantly by human researchers and conventional scanners, was therefore set \emph{before} the capability shift its obligations must now govern, its threat assumptions notarised barely ten weeks before the technology that would void them went public. Put plainly, the \cra{} switches on its central obligations in December~2027 calibrated to a threat model the field had already abandoned years earlier. We advance this as a European organisation working on Cybersecurity AI, and we advance it constructively: the \cra{} is not wrong so much as early to its own obsolescence, and our purpose in diagnosing that while there is still time to act is to help Europe avert a fate that is not yet sealed.

\paragraph{Scope and terminology.} Two questions about AI and the \cra{} must be kept apart. The security \emph{of} AI systems treated as products, whether prompt injection or goal hijacking fits the \cra{}'s obligations, is a real question but not ours. Our subject is the second: AI agents used \emph{as instruments of security}, \cai{} systems that discover, weaponise, and exploit vulnerabilities in \emph{other} products, and how, by transforming the adversarial landscape, they invalidate the assumptions on which the \cra{}'s process rests. Throughout, ``agent'' means an offensive or defensive security agent acting on the ecosystem, not an AI product being regulated. The distinction sharpens the claim: the \cra{} can be defeated without any AI system ever being the regulated product. We also write ``automated'' rather than ``autonomous,'' because today's leading agents operate at partial autonomy with humans in the loop~\citep{mayoral2025autonomy}, and the remedy we propose is a human-\cai{} partnership, not a replacement of it.

\paragraph{The premises and the thesis.} A process-oriented regime does not certify that a product is secure; it certifies that a manufacturer ran a defined process. That certification is meaningful only if four background premises about the vulnerability lifecycle hold (Section~\ref{sec:premises}): \premise{1}~discovery is human-scarce; \premiseb{2}~a product's set of known exploitable vulnerabilities is knowable at market placement; \premiseb{3}~exploitation is a discrete, detectable event; and \premiseb{4}~remediation keeps pace with discovery. \cai{} agents stress all four, and the \cra{} responds in two opposite ways (Figure~\ref{fig:hero}). Against the sheer \emph{volume} agents surface, the regime \bends{}: \premise{1} is strained, and a risk-based process re-centres from ``did you remediate everything?'' toward ``did you prioritise defensibly, and document it?'' Against the \emph{collapse} of the lifecycle's tempo and economics, the regime \breaks{}: agents withdraw \premiseb{2}\,\premiseb{3}\,\premiseb{4}, so a conformant product becomes exploitable with no change to it, and a point-in-time certificate attests to a posture that no longer holds. The mismatch is with the landscape, not the product, so no amount of process quality restores validity. This is not a plea for patience: the capability that voids these premises is deployed, measured, and cheap \emph{today}, while the obligations that ignore it do not fully bind until December~2027, so a regulator or a manufacturer that treats the agentic shift as a future contingency is already behind it. The consequence runs straight through the paper: security has to move from a periodic, human-run process to a continuous, agent-operated one, and that is precisely the move a credible \cra{} now demands of itself.

\paragraph{Contributions and related work.} We reconstruct the \cra{}'s process orientation and its four load-bearing premises from the enacted text; map each mechanism to the \cai{}-agent dynamic that stresses it, with a verdict per provision (Table~\ref{tab:mechmap}); state six falsifiable predictions through 2028; propose an agent-native remedy, continuous and agent-operated conformity, grounded in the same AI-security research that documents the threat; and validate that remedy on two \cra{}-scope robots, a humanoid and a robotic lawn mower, where a \cai{} defender (the \href{https://aliasrobotics.com/ris.php}{Robot Immune System}) contains attacks that fully compromise the undefended, conformant systems (Section~\ref{sec:solution}). The analysis joins three literatures usually kept apart. Legal scholarship characterises the \cra{} as hybrid, self-assessment-default governance~\citep{teichmann2025cra} and dissects its reporting logic, contrasting the \emph{actively exploited} trigger with the US \emph{known exploited} construct~\citep{ruohonen2025cra}, while a parallel critique argues the manufacturer-centric model fits open-source software poorly~\citep{linuxfoundation2023cra}. A fast-growing empirical literature documents agents discovering and exploiting vulnerabilities~\citep{fang2024oneday,fang2024zeroday,bigsleep2024,darpa_aixcc2025,csa_exploit2026}, and operational \cai{} frameworks report large speed and cost advantages while showing the attacker edge is conditional~\citep{cai2025,cai_attackdefense2025,cai_robots2026,cai_synthetic_apt2026}. A third strand models the vulnerability lifecycle economically, from Akerlof's market for lemons to the disclosure-to-exploit window now measured collapsing toward zero~\citep{akerlof1970,anderson2006economics,zerodayclock2026}. We read the first against the second and third. To our knowledge this premise-level bends-versus-breaks decomposition, tied provision-by-provision to the enacted text, is new. Every legal claim is anchored to an article or annex of the Regulation; every capability claim to a dated primary source.

%% file: tex/regime.tex
We reconstruct the \cra{}'s process orientation from three features of the enacted text, state the four premises that orientation silently assumes, and show that the interval between assumption and obligation is exactly the interval in which the vulnerability ecosystem inverted.

\subsection{A process, not an outcome}
Three features make the \cra{} process-oriented rather than outcome-oriented. \emph{First, a thin, risk-conditioned property floor.} Annex~I, Part~I lists product properties, but the operative ones are gated on a risk assessment. Even the strongest-sounding requirement, that products be \emph{``made available on the market without known exploitable vulnerabilities''} (Annex~I, Part~I, (2)(a)), is expressly conditioned on the Article~13(2) risk assessment. It is a floor against \emph{known} exploitable flaws at market placement, not a guarantee of security, and Article~3 fixes its scope: a \emph{vulnerability} can be exploited by a cyber threat (Art.~3(40)), an \emph{exploitable vulnerability} one an adversary could effectively use under practical conditions (Art.~3(41)). \emph{Second, substantive obligations are handling processes.} Annex~I, Part~II reads as a process specification: identify and document vulnerabilities, produce an \textsc{sbom}, ``address and remediate vulnerabilities without delay,'' test regularly, disclose fixed flaws, run coordinated disclosure, and distribute updates securely. Article~13 frames the duty as performing and \emph{documenting} a risk assessment ``updated as appropriate during a support period'' and exercising due diligence over third-party components. The verbs are process verbs (assess, document, handle, disclose, update), not outcome verbs. \emph{Third, reporting is triggered by exploitation, not existence.} Article~14 requires notifying only an \emph{``actively exploited vulnerability,''} one for which there is reliable evidence a malicious actor has exploited it (Art.~3(42)), on a 24-hour, 72-hour, 14-day cadence through the ENISA platform. A newly discovered bug is generally not reportable. One further provision fixes scope: conformity is assessed once, against the product's state at placing (Article~32), and re-triggered only by a \emph{``substantial modification''} of the product (Art.~3). Conformity is thus a point-in-time gate on a static artefact, placing the \cra{} much closer to continuous cyber-risk management than to a zero-defect mandate.

\subsection{The four premises of the process model}\label{sec:premises}
The conditional promise of a process-oriented regime, run the process and the product is acceptably secure, holds only where the environment behaves as the process assumes. Four premises carry the weight of the whole edifice, and they are nowhere stated in the Regulation, which is exactly the point: they are the load-bearing walls no one thought to draw on the blueprint, the tacit model of the world the drafters inhabited.
\begin{description}[leftmargin=1.4em,itemsep=3pt,topsep=3pt]
\item[\premise{1}~\textnormal{\textit{Discovery is human-scarce.}}] Finding an exploitable vulnerability takes costly, skilled human effort, so what is \emph{found} is a small, slowly growing subset of what is latent. Article~14's ``report only what is actively exploited'' and Annex~I's ``handle what you find'' both assume the found set is manageable because discovery is expensive.
\item[\premiseb{2}~\textnormal{\textit{Posture is knowable at a point in time.}}] A product's set of known exploitable vulnerabilities can be enumerated at placement, making the Annex~I, Part~I, (2)(a) condition and the Article~32 assessment meaningful as attestations. The premise is that ``known'' at time $t$ approximates ``knowable'' shortly after $t$.
\item[\premiseb{3}~\textnormal{\textit{Exploitation is a discrete, detectable event.}}] Active exploitation is rare and distinct enough that ``reliable evidence'' of it (Art.~3(42)) is an informative signal worth a 24/72-hour duty. The premise is that exploitation events are countable and separable from background noise.
\item[\premiseb{4}~\textnormal{\textit{Remediation keeps pace with discovery.}}] The interval between a vulnerability becoming known and its being weaponised is long enough for ``remediate without delay'' to be a winnable race, so a support period of scheduled updates is an adequate control.
\end{description}
\noindent \cai{} agents weaken \premise{1} \emph{quantitatively} (the regime bends, Section~\ref{sec:bends}) and falsify \premiseb{2}\,\premiseb{3}\,\premiseb{4} \emph{environmentally} (the regime breaks, Section~\ref{sec:breaks}). A strained premise leaves the promise intact but harder to satisfy; a withdrawn premise voids it, because the certificate now attests to a state of the world the agent-transformed landscape no longer contains.

\subsection{The assumption-formation window}\label{sec:timing}
\begin{figure*}[t]
\centering
\resizebox{0.99\textwidth}{!}{\input{tex/fig_timeline.tex}}
\caption{\textbf{The assumption-formation window.} The \cra{} legislative track (top, deep teal-blue) against the Cybersecurity AI (\cai{}) agent-capability track (bottom), each milestone cited in place. Markers carry three classes: \textcolor{cap_color}{\textbf{red}} for general, published capability; \textcolor{alias_teal}{\textbf{teal}} for the Alias Robotics \cai{} lineage used \emph{offensively}, with the open-source \cai{} framework itself as the flagship (star, \citealp{cai2025}); and \textcolor{cai_accent}{\textbf{green}} for the \emph{same} lineage turned to \emph{defence}, the empirical basis of the remedy in Section~\ref{sec:solution} (agentic defenders out-patching attackers, and a self-hostable on-premise defender containing a frontier attacker). The shaded band marks the pre-ChatGPT interval in which the \cra{}'s assumptions were formed; each offensive event is tagged with the premise (P1--P4, Section~\ref{sec:premises}) it undermines. By the time Article~14 binds (September~2026) and the regime fully applies (December~2027), \cai{} agents had already exploited real one-day CVEs (April~2024), rediscovered latent flaws on demand (November~2024), discovered vulnerabilities at population scale and commodity cost (DARPA AIxCC, August~2025), and compromised \cra{}-scope consumer robots (March~2026).}
\label{fig:timeline}
\end{figure*}
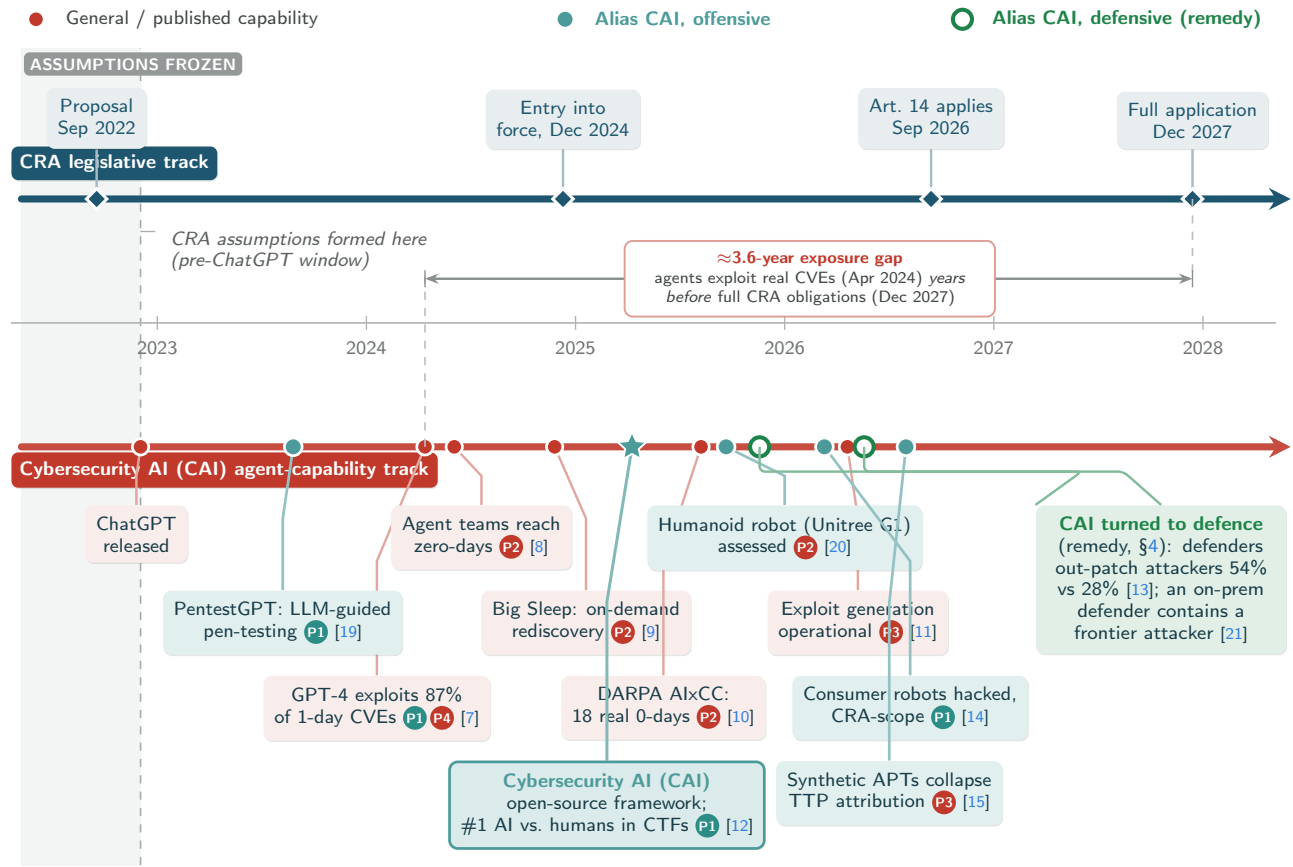

The \cra{}'s obligations bind late relative to the capability shift they must govern, and Figure~\ref{fig:timeline} places the two tracks side by side. The gap is not a projection but a measurement: every point on the lower track is a published, dated result, and the most recent, \cai{} agents compromising \cra{}-scope consumer products, predates the Regulation's own headline obligations. A full-text reading of the enacted Regulation confirms the blind spot. It mentions ``artificial intelligence'' only to cross-reference the AI Act, contains no occurrence of ``machine learning,'' and nowhere treats AI as a \emph{capability of the adversary}. For a cybersecurity statute finalised in 2024, that is a loud silence: the adversary's most consequential new recruit goes entirely unnamed. The point-in-time conformity it certifies is therefore blind to the one variable, adversary tooling, that the lower track is changing fastest.

The datable inflection points, restricted to agents acting as security instruments \emph{on other products}, run from \emph{PentestGPT} (2023, \textsc{cai} lineage) demonstrating LLM-guided testing (P1)~\citep{pentestgpt2024}; through a single GPT-4 agent exploiting 87\% of 15 real one-day CVEs from their descriptions, versus 0\% for scanners and open models (April~2024)~\citep{fang2024oneday}, weakening \premise{1} and near-eliminating the disclosure-to-weaponisation gap (P4); the open \emph{Cybersecurity AI} framework ranking as the top AI team against humans in global CTFs at up to two orders of magnitude lower cost (2025)~\citep{cai2025}; planner-plus-subagent teams reaching real zero-days~\citep{fang2024zeroday}; Google's Big Sleep rediscovering a latent memory-safety flaw in SQLite on demand (November~2024, P2)~\citep{bigsleep2024}; the DARPA AIxCC final identifying 86\% of synthetic flaws and 18 real zero-days across 54~million lines of code at roughly \$152 per task (August~2025)~\citep{darpa_aixcc2025}; agent-emulated synthetic APTs collapsing TTP-based attribution (2026, P3)~\citep{cai_synthetic_apt2026}; a \cai{} agent operationalised on a production humanoid robot (Unitree G1, September~2025, P2)~\citep{cai_humanoid2025}; and an open-source agent compromising three consumer robots, finding fleet-wide flaws and showing attacks once needing deep expertise can now be run ``by anyone'' (March~2026)~\citep{cai_robots2026}, which withdraws the specialised-attacker-scarcity premise (P1) for the exact product class the Regulation governs. Crucially, the same \textsc{cai} lineage also produced the \emph{defensive} results that ground the remedy (Section~\ref{sec:solution}): in controlled Attack/Defense evaluations agentic defenders out-patch attackers~\citep{cai_attackdefense2025}, and a self-hostable on-premise defender contains a frontier-class attacker~\citep{cai_dynamicranges2026}. The scarce resource in vulnerability management has moved from \emph{finding} flaws to \emph{deciding which matter} (the bend), while the stability of a certified posture has evaporated (the break). The next section separates the two.

%% file: tex/fig_timeline.tex
\providecommand{\tpb}[2]{\tikz[baseline=(tb.base)]{\node[circle, fill=#1, text=white,
   font=\tiny\bfseries\sffamily, inner sep=0.5pt, minimum size=1.0em](tb){#2};}}
\begingroup\renewcommand{\cra}{CRA}\renewcommand{\cai}{CAI}%
\begin{tikzpicture}[
    every node/.style={font=\footnotesize\sffamily, color=cai_dark},
    xscale=2.7, yscale=1.0,
    cardshadow/.style={drop shadow={shadow xshift=0.3pt, shadow yshift=-0.5pt,
        opacity=0.14, fill=cai_dark}},
    cracard/.style={rounded corners=3pt, fill=cra_color!10, text=cra_color,
        align=center, inner sep=4pt, font=\footnotesize\sffamily, cardshadow},
    capcard/.style={rounded corners=3pt, fill=cap_color!9, text=human_color,
        align=center, inner sep=4pt, font=\footnotesize\sffamily, cardshadow},
    caicard/.style={rounded corners=3pt, fill=alias_teal!16, text=human_color,
        align=center, inner sep=4pt, font=\footnotesize\sffamily, cardshadow},
    caicardF/.style={rounded corners=3pt, fill=alias_teal!20, draw=alias_teal, line width=1pt,
        text=human_color, align=center, inner sep=4pt, font=\footnotesize\sffamily, cardshadow},
    caicardD/.style={rounded corners=3pt, fill=cai_accent!13, text=human_color,
        align=center, inner sep=4pt, text width=2.95cm, font=\footnotesize\sffamily, cardshadow},
    dotC/.style={diamond, fill=cra_color, draw=white, line width=1.1pt, inner sep=0pt, minimum size=8pt},
    dotK/.style={circle, fill=cap_color, draw=white, line width=1pt, inner sep=0pt, minimum size=6pt},
    dotA/.style={circle, fill=alias_teal, draw=white, line width=1pt, inner sep=0pt, minimum size=6.5pt},
    dotF/.style={star, star points=5, star point ratio=2.1, fill=alias_teal, draw=white,
        line width=0.7pt, inner sep=0pt, minimum size=11pt},
    dotD/.style={circle, fill=white, draw=cai_accent, line width=1.5pt, inner sep=0pt, minimum size=7pt},
]
\def\yA{1.6}   
\def\yB{-1.6}  

\fill[cai_dark!6] (0.35,-7.0) rectangle (0.92,3.55);
\draw[cai_dark!45, dashed, line width=0.6pt] (0.92,-7.0) -- (0.92,3.55);
\node[fill=cai_dark!55, text=white, rounded corners=2pt, inner sep=2.2pt, font=\scriptsize\bfseries\sffamily,
      anchor=north west] at (0.36,3.5) {ASSUMPTIONS FROZEN};
\node[align=left, color=cai_dark!85, font=\footnotesize\itshape\sffamily, anchor=north west] at (1.02,1.30)
  {\cra{} assumptions formed here\\(pre-ChatGPT window)};
\draw[cai_dark!45, line width=0.5pt] (0.99,1.18) -- (0.92,1.18);

\draw[cai_dark!40, line width=0.6pt] (0.30,0) -- (6.35,0);
\foreach \yr in {2023,2024,2025,2026,2027,2028}{
  \pgfmathsetmacro\xx{\yr-2022}
  \draw[cai_dark!40, line width=0.5pt] (\xx,-0.09) -- (\xx,0.09);
  \node[anchor=north, color=cai_dark!75, font=\footnotesize\sffamily] at (\xx,-0.13) {\yr};
}

\draw[cra_color, line width=2.6pt, line cap=round, -{Stealth[length=8pt,width=8pt]}]
  (0.35,\yA) -- (6.42,\yA);
\draw[cap_color!90, line width=2.6pt, line cap=round, -{Stealth[length=8pt,width=8pt]}]
  (0.35,\yB) -- (6.42,\yB);
\node[anchor=west, fill=cra_color, text=white, rounded corners=2.5pt, inner sep=3pt,
      font=\footnotesize\bfseries\sffamily, cardshadow] at (0.30,\yA+0.46) {\cra{} legislative track};
\node[anchor=west, fill=cap_color, text=white, rounded corners=2.5pt, inner sep=3pt,
      font=\footnotesize\bfseries\sffamily, cardshadow] at (0.30,\yB-0.30) {Cybersecurity AI (\cai{}) agent-capability track};

\foreach \x/\h/\lab in {0.71/0.64/{Proposal\\Sep 2022}, 2.94/0.64/{Entry into\\force, Dec 2024}, 4.70/0.64/{Art.~14 applies\\Sep 2026}, 5.95/0.64/{Full application\\Dec 2027}}{
  \draw[cra_color!40, line width=0.9pt] (\x,\yA) -- (\x,\yA+\h);
  \node[cracard, anchor=south] at (\x,\yA+\h) {\lab};
  \node[dotC] at (\x,\yA) {};
}

\node[dotK] at (0.42,3.92) {};
\node[anchor=west, color=cai_dark, font=\footnotesize\sffamily] at (0.52,3.92) {General / published capability};
\node[dotA] at (2.95,3.92) {};
\node[anchor=west, color=alias_teal, font=\footnotesize\bfseries\sffamily] at (3.05,3.92) {Alias \cai{}, offensive};
\node[dotD] at (4.85,3.92) {};
\node[anchor=west, color=cai_accent, font=\footnotesize\bfseries\sffamily] at (4.95,3.92) {Alias \cai{}, defensive (remedy)};

\foreach \x/\h/\dx/\lab in {%
  0.92/0.75/-0.02/{ChatGPT\\released},%
  2.28/2.95/-0.23/{GPT-4 exploits 87\%\\of 1-day CVEs \tpb{bends_color}{P1}\,\tpb{breaks_color}{P4}~{\scriptsize\citep{fang2024oneday}}},%
  2.42/0.75/0.13/{Agent teams reach\\zero-days \tpb{breaks_color}{P2}~{\scriptsize\citep{fang2024zeroday}}},%
  2.90/1.85/0.15/{Big Sleep: on-demand\\rediscovery \tpb{breaks_color}{P2}~{\scriptsize\citep{bigsleep2024}}},%
  3.60/2.95/-0.18/{DARPA AIxCC:\\18 real 0-days \tpb{breaks_color}{P2}~{\scriptsize\citep{darpa_aixcc2025}}},%
  4.30/1.85/0.05/{Exploit generation\\operational \tpb{breaks_color}{P3}~{\scriptsize\citep{csa_exploit2026}}}%
}{
  \draw[cap_color!45, line width=0.9pt] (\x,\yB) -- (\x+\dx,\yB-\h*0.5) -- (\x+\dx,\yB-\h);
  \node[capcard, anchor=north] at (\x+\dx,\yB-\h) {\lab};
  \node[dotK] at (\x,\yB) {};
}

\foreach \x/\h/\dx/\lab in {%
  1.65/1.85/-0.05/{PentestGPT: LLM-guided\\pen-testing \tpb{bends_color}{P1}~{\scriptsize\citep{pentestgpt2024}}},%
  3.72/0.75/0.28/{Humanoid robot (Unitree G1)\\assessed \tpb{breaks_color}{P2}~{\scriptsize\citep{cai_humanoid2025}}},%
  4.19/2.95/0.41/{Consumer robots hacked,\\\cra{}-scope \tpb{bends_color}{P1}~{\scriptsize\citep{cai_robots2026}}},%
  4.58/4.05/-0.08/{Synthetic APTs collapse\\TTP attribution \tpb{breaks_color}{P3}~{\scriptsize\citep{cai_synthetic_apt2026}}}%
}{
  \draw[alias_teal!60, line width=1.0pt] (\x,\yB) -- (\x+\dx,\yB-\h*0.5) -- (\x+\dx,\yB-\h);
  \node[caicard, anchor=north] at (\x+\dx,\yB-\h) {\lab};
  \node[dotA] at (\x,\yB) {};
}
\draw[alias_teal!70, line width=1.1pt] (3.27,\yB) -- (3.15,\yB-2.0) -- (3.15,\yB-4.05);
\node[caicardF, anchor=north] at (3.15,\yB-4.05)
  {\textbf{\textcolor{alias_teal}{Cybersecurity AI (\cai{})}}\\open-source framework;\\\#1 AI vs.\ humans in CTFs \tpb{bends_color}{P1}~{\scriptsize\citep{cai2025}}};
\node[dotF] at (3.27,\yB) {};

\node[dotD] at (3.88,\yB) {};
\node[dotD] at (4.38,\yB) {};
\node[caicardD, anchor=north] (gdef) at (5.80,\yB-0.75)
  {\textbf{\textcolor{cai_accent}{\cai{} turned to defence}} (remedy, \S\ref{sec:solution}): defenders out-patch attackers 54\% vs 28\%~{\scriptsize\citep{cai_attackdefense2025}}; an on-prem defender contains a frontier attacker~{\scriptsize\citep{cai_dynamicranges2026}}};
\draw[cai_accent!55, line width=0.9pt, rounded corners=2pt] (3.88,\yB) -- (3.88,\yB-0.32) -- (5.42,\yB-0.32) -- (gdef.north west);
\draw[cai_accent!55, line width=0.9pt, rounded corners=2pt] (4.38,\yB) -- (4.38,\yB-0.32) -- (5.62,\yB-0.32) -- (gdef.north);

\draw[cai_dark!35, dashed, line width=0.6pt] (2.28,\yB) -- (2.28,0.60);
\draw[cai_dark!35, dashed, line width=0.6pt] (5.95,\yA) -- (5.95,0.60);
\draw[cai_dark!60, line width=0.5pt] (2.28,0.46) -- (2.28,0.68);
\draw[cai_dark!60, line width=0.5pt] (5.95,0.46) -- (5.95,0.68);
\draw[{Stealth[length=5pt]}-{Stealth[length=5pt]}, cai_dark!60, line width=0.9pt] (2.28,0.57) -- (5.95,0.57);
\node[fill=white, draw=cap_color!50, line width=0.7pt, rounded corners=3pt, inner sep=4pt,
      font=\scriptsize\sffamily, align=center, text=cai_dark, text width=4.5cm] at (4.12,0.57)
  {{\bfseries\color{cap_color}$\approx$3.6-year exposure gap}\\
   agents exploit real CVEs (Apr 2024) \emph{years before} full \cra{} obligations (Dec 2027)};
\end{tikzpicture}\endgroup

%% file: tex/core.tex
\cai{} agents stress the \cra{} in two ways that call for opposite responses. Volume is the \emph{quantitative} face: it strains \premise{1} but leaves the process model's internal logic intact, so the regime \bends{}. The collapse of the lifecycle's tempo and economics is the \emph{environmental} face: it withdraws \premiseb{2}\,\premiseb{3}\,\premiseb{4}, so the regime \breaks{}. This section separates the two, maps every load-bearing mechanism to the dynamic that stresses it (Table~\ref{tab:mechmap}), and states what each response predicts.

\subsection{The regime bends: vulnerability abundance}\label{sec:bends}
Abundance weakens \premise{1} (discovery is human-scarce) without withdrawing it, and it is absorbable because the process model never assumed abundance would not occur, only that \emph{attention} would be scarce, which abundance intensifies rather than negates. When security agents surface many candidate vulnerabilities, the triage queue expands: the 2026 Cloud Security Alliance synthesis frames automated exploit generation as having crossed from demonstration into operational tooling~\citep{csa_exploit2026}, and the pressure is on throughput, more candidates per unit time than any human process was provisioned for.

The honest evidence base is more sober than the headlines, and the conservative reading \emph{strengthens} the argument. The widely cited 87\% one-day result~\citep{fang2024oneday} holds only when the agent is handed the CVE description; the same study reports 7\% without it. On harder real-world benchmarks, the best frameworks resolve up to 13\% on \textsc{cve-bench}~\citep{cvebench2025} and about 20\% on \textsc{cybergym} across 1{,}507 vulnerabilities~\citep{cybergym2025}. Yet even at that rate the same \textsc{cybergym} run surfaced 34 previously unknown zero-days and 18 incomplete historical patches~\citep{cybergym2025}, and Meta's \textsc{cyberseceval}~3 already benchmarks offence as a first-class model capability~\citep{cyberseceval3}. Abundance does not require superhuman agents, only cheap and parallel ones: a tool that finds one real flaw in five, run continuously across a dependency graph, still overruns a human handling process.

Nothing in the \cra{} requires zero vulnerabilities or complete remediation; Article~14's exploitation trigger and Article~13's risk-based framing were built for a world in which not every flaw can be chased. Faced with more candidates, the process runs against a larger input without contradicting itself, and in practice compliance re-centres from \emph{remediation-completeness} toward \emph{defensible prioritisation}. That prioritisation is itself getting harder to do well: on the ENISA European Vulnerability Database, the predictors it would lean on lose discriminating power, with EPSS ranking and CVSS severity decoupling from real exploitation tempo (Appendix~\ref{app:euroclock}). The questions an authority (Article~52) will turn on are already in the model: was exploitation actually occurring (Art.~3(42)), was the flaw reachable in the shipped configuration (Art.~13(2)), was an effective mitigation available, and was the prioritisation reasonable and documented (Annex~I, Part~II)? This mirrors the move mature safety engineering made long ago. The Article~14 trigger, the 24/72/14 cadence, coordinated disclosure, and the risk-based handling process all bend: strained but not contradicted. Two strains bear watching, and we treat them as predictions rather than breaks: the reporting cadence may prove too slow against machine-speed weaponisation, and coordinated disclosure weakens when a single technique generalises across a whole device family at once.

\subsection{The regime breaks: landscape collapse}\label{sec:breaks}
The second stressor is where deference to the \cra{}'s craftsmanship has to stop. This is not a process overwhelmed with work; it is a process whose \emph{world has ceased to exist}, so that a manufacturer can run it flawlessly, fund it fully, and pass every audit, and still ship a certificate that is false the week after it is signed. \cai{} agents force this by changing the tempo, economics, and symmetry of the lifecycle, withdrawing \premiseb{2}\,\premiseb{3}\,\premiseb{4}. When a premise is withdrawn the conformity the \cra{} issues does not become harder to earn; it becomes \emph{invalid}, attesting to a posture the landscape no longer sustains.

\begin{definition}[Landscape validity]
A point-in-time conformity attestation is \emph{landscape-valid} if the posture it certifies at placement remains a good approximation of the product's posture over the period it is relied upon. A mechanism \bends{} when validity is strained but recoverable through prioritisation; it \breaks{} when the agent-transformed landscape withdraws the premise that made the attestation meaningful, so that re-running the process cannot restore it.
\end{definition}

\paragraph{Certified-secure becomes trivially-exploitable, product unchanged (P2).} Annex~I, Part~I, (2)(a) requires placement \emph{``without known exploitable vulnerabilities.''} ``Known'' is a point-in-time predicate presuming posture is stable shortly after placement. \cai{} discovery falsifies it: a latent flaw unknown at assessment can be rediscovered on demand the next week at order \$100 per attempt~\citep{fang2024oneday,darpa_aixcc2025}. The product has not changed, nor its declared conformity, yet its real exploitable set has expanded because the \emph{environment's} capacity to enumerate it has. The certificate attests to a property that is technically true and practically empty, because ``known'' has decoupled from ``findable.'' It has not lied so much as answered, with great precision, a question the world has stopped asking, and the next assessment is only another snapshot against a moving environment.

\paragraph{Exploitation stops being a discrete signal (P3).} Article~14 reports only actively-exploited vulnerabilities (Art.~3(42)), a design presuming exploitation is rare and detectable enough to be an informative trigger. When offensive tooling is automated and continuous~\citep{csa_exploit2026}, exploitation becomes background rather than event: ``reliable evidence that a malicious actor has exploited'' the flaw is either everywhere (every exposed product is probed continuously) or nowhere (agent-driven exploitation leaves a fainter human signature). Either way a duty keyed to a rare discrete event does not scale to a regime where the event is ambient.

\paragraph{``Remediate without delay'' describes an unwinnable race (P4).} Annex~I, Part~II(2) is a control only if the disclosure-to-weaponisation window exceeds the update cycle. That window was \emph{already} closing before agents arrived, which is why the break is structural, not speculative. Aggregating 3{,}549 CVE-exploit pairs across ten sources, the median time-to-exploit fell from 771~days in 2018 to 68 in 2021, 5.3 in 2023, and zero in 2025, an exponential-decay fit at $R^2=0.98$ reaching a one-day median by 2028~\citep{zerodayclock2026}. Over the same span the share weaponised at or before disclosure rose from 19\% to 54\%, and within 24~hours from 19\% to 58\% (Appendix~\ref{app:zerodayclock}); peer-reviewed open-source data corroborates the direction~\citep{oss_disclosure2025}, as does industry telemetry showing exploitation now often preceding patch availability~\citep{mandiant2024tte}, and Europe's own registry reproduces it: the ENISA European Vulnerability Database shows the same collapse toward a near-zero median (Appendix~\ref{app:euvd}). The same capability that let an agent weaponise a one-day CVE from its description alone~\citep{fang2024oneday} pushes this trend past the point where a scheduled-update model can keep pace. Against a weaponisation window now measured in hours, a scheduled patch cadence brings a calendar to a stopwatch's fight, and running the process better does not close a gap defined by the adversary's tempo, not the manufacturer's diligence.

\paragraph{Point-in-time conformity certifies a vanishing instant (P2--P4 jointly).} These breaks share a root: Article~32 conformity is assessed once and re-triggered only by a \emph{``substantial modification''} the manufacturer makes to the product (Art.~3). But the events that invalidate the attestation now originate in the \emph{environment}: a new open-source offensive agent, a drop in the cost of discovery, a technique that generalises across a device family. None is a modification of any product, so none re-triggers assessment, yet each can move a whole population of conformant products from secure to exploitable overnight. The gate has no sensor for the variable that now dominates risk.

\subsubsection*{Why this is a break, not a bend}
A bend leaves the conditional promise intact but costlier, and the \cra{}'s own flexibility supplies the remedy. A break withdraws a premise, so that even a perfectly executed process issues a certificate that is false in the world the agents create. The economic stakes are clearest in the CE mark's function. Information-security markets are a textbook case of Akerlof's market for lemons~\citep{akerlof1970}: buyers cannot observe security, so absent a credible signal the market prices every product as insecure and investment collapses~\citep{anderson2006economics}. The \cra{}'s conformity regime supplies that missing signal, a legitimate and valuable goal. But a signal helps only while it correlates with quality. When agents withdraw \premiseb{2}--\premiseb{4}, the attestation is still \emph{issued} on the same terms while the property it certifies evaporates, so the signal decouples from reality, and a decoupled safety signal is worse than none: it transfers risk to the buyers and integrators least able to see it while discharging the obligation on paper. A false mark carrying the authority of Union law is not a neutral failure but an active harm: it manufactures the very market-for-lemons assurance the \cra{} was enacted to abolish, now underwritten by the state. A market-for-lemons regime is meant to drive the bad cars off the lot; decoupled, this one staples a certificate of freshness to every lemon it was built to expose. Crucially, this is \emph{independent of whether the regulated product contains any AI at all}. A conventional IP camera, firmware frozen and fully conformant, is moved from secure to exploitable by an offensive agent operating around it. That is why the problem cannot be delegated to the AI Act: the object that changed is not the product but the adversary, and the \cra{} has no mechanism that takes the adversary's evolving capability as an input.

\subsection{A mechanism-level stress map}\label{sec:mechmap}
Table~\ref{tab:mechmap} makes the split concrete, mapping each load-bearing \cra{} mechanism to the \cai{}-agent dynamic that stresses it, the premise it depends on, and a verdict, ordered from most absorbable (top) to structural break (bottom). Read as a verdict sheet the table is unflattering: the provisions that survive are the ones that merely help the process cope, while every mechanism that makes the certificate \emph{mean} something, the market-entry condition, the actively-exploited trigger, and the point-in-time gate, is scored a break. The pattern is systematic: mechanisms resting only on \premise{1} \bends{}, because \premise{1} is strained but not withdrawn; mechanisms resting on \premiseb{2}--\premiseb{4} \breaks{}, because agents remove a feature of the landscape (stable posture, rare exploitation, a winnable remediation race) that the mechanism silently relies on. The two intermediate rows are premises mid-withdrawal, the provisions where guidance-level fixes will first prove insufficient and the leading indicators for the breaks predictions below.

\begin{table*}[t]
\centering
\small
\setlength{\tabcolsep}{5pt}
\renewcommand{\arraystretch}{1.35}
\caption{\textbf{Mechanism-level stress map.} Each \cra{} mechanism against the \cai{}-agent dynamic that stresses it, the lifecycle premise it relies on, and a verdict. Verdicts: \textcolor{bends_color}{\textbf{Bends}} (absorbable by re-interpreting existing obligations); \textcolor{mid_color}{\textbf{Bends$\rightarrow$Breaks}} (currently absorbable, premise eroding); \textcolor{breaks_color}{\textbf{Breaks}} (premise withdrawn, a new construct required). Premises: P1 discovery human-scarce; P2 posture knowable at a point in time; P3 exploitation discrete/detectable; P4 remediation keeps pace.}
\label{tab:mechmap}
\small
\setlength{\tabcolsep}{5pt}
\begin{tabular}{p{4.0cm}p{2.2cm}p{5.9cm}>{\raggedright\arraybackslash}p{1.05cm}>{\raggedright\arraybackslash}p{2.4cm}}
\toprule
\rowcolor{cra_color!13}
\textbf{\cra{} mechanism} & \textbf{Citation} & \textbf{\cai{}-agent dynamic} & \textbf{Prem.} & \textbf{Verdict}\\
\midrule
Risk-based, process-oriented handling & Art.~13(2)--(3); Annex~I Pt~II & Agents surface more candidates; triage becomes the scarce resource, the design's home ground & P1 & \cellcolor{bends_color!16}\textcolor{bends_color}{\textbf{Bends}}\\
24\,h / 72\,h / 14-day reporting cadence & Art.~14(2) & Machine-speed weaponisation compresses time-to-impact but events remain reportable & P1 & \cellcolor{bends_color!16}\textcolor{bends_color}{\textbf{Bends}}\\
Coordinated vulnerability disclosure & Annex~I Pt~II(5) & One agent-found technique generalises across a device family; the per-product model strains & P1,P2 & \cellcolor{mid_color!20}\textcolor{mid_color}{\textbf{Bends$\rightarrow$Breaks}}\\
Report only actively-exploited vulns & Art.~14; 3(42) & Automated, continuous exploitation makes the ``rare discrete event'' trigger uninformative & P3 & \cellcolor{breaks_color!16}\textcolor{breaks_color}{\textbf{Breaks}}\\
``No known exploitable vulns'' at placing & Annex~I Pt~I, (2)(a) & On-demand rediscovery decouples ``known'' from ``findable''; certified-secure becomes exploitable, product unchanged & P2 & \cellcolor{breaks_color!16}\textcolor{breaks_color}{\textbf{Breaks}}\\
Patch-based ``remediate without delay'' & Annex~I Pt~II(2) & Disclosure-to-weaponisation window closes below the fleet update cycle & P4 & \cellcolor{breaks_color!16}\textcolor{breaks_color}{\textbf{Breaks}}\\
Point-in-time conformity assessment & Art.~32; Annex~I~Pt~I & Certifies a landscape snapshot; the risk-dominant variable now lives in the environment & P2 & \cellcolor{breaks_color!16}\textcolor{breaks_color}{\textbf{Breaks}}\\
Re-conformity on ``substantial modification'' & Art.~3; Art.~32 & Invalidating events (new offensive agent, cheaper discovery) are not product modifications, so no trigger fires & P2 & \cellcolor{breaks_color!16}\textcolor{breaks_color}{\textbf{Breaks}}\\
Support-period lifecycle obligation & Art.~13(8); Art.~3 & Support keyed to product lifetime, not to adversary-capability growth that outpaces it & P4 & \cellcolor{mid_color!20}\textcolor{mid_color}{\textbf{Bends$\rightarrow$Breaks}}\\
\bottomrule
\end{tabular}
\end{table*}

\subsection{Predictions}\label{sec:predictions}
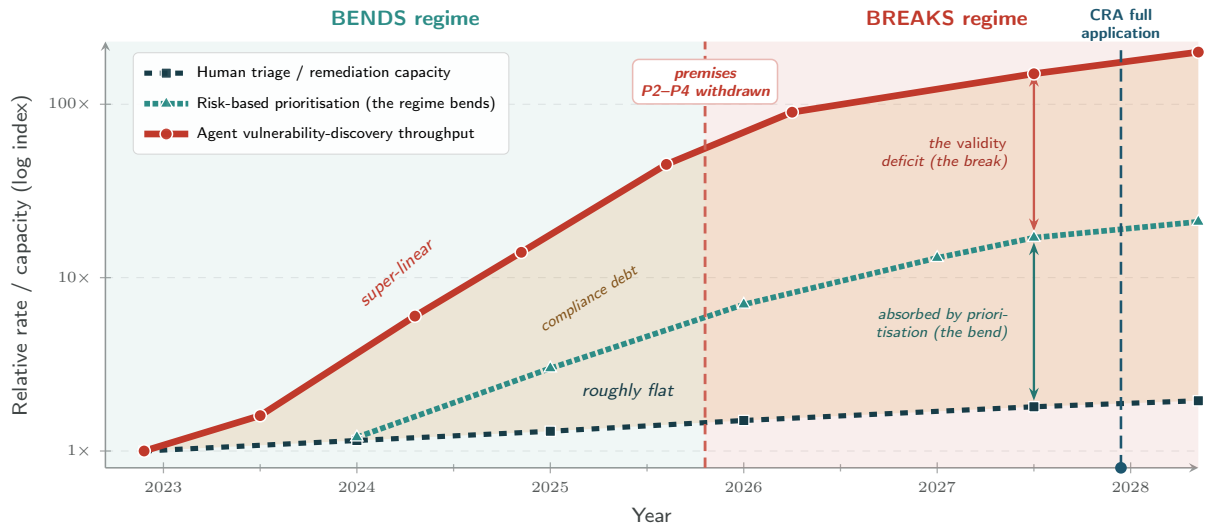
\begin{figure*}[t]
\centering
\resizebox{0.92\textwidth}{!}{\input{tex/fig_predictions.tex}}
\caption{\textbf{Projected divergence and the regime transition (schematic).} Agent-driven vulnerability-discovery throughput (red) rises super-linearly; human triage/remediation capacity (navy, dashed) stays roughly flat; risk-based prioritisation (teal, dotted) partly closes the gap, the \cra{} bending (P1 strained). The vertical boundary marks where \cai{} agents withdraw P2--P4 (stable posture, discrete exploitation, winnable remediation), beyond which the mechanisms in Table~\ref{tab:mechmap} enter the \scw{breaks} regime: a validity deficit no throughput gain can address. Curves are illustrative of the argued dynamic, not fitted measurements; the log axis is a unitless index normalised to the 2022 human baseline.}
\label{fig:predictions}
\end{figure*}

The decomposition yields falsifiable expectations about how the regime behaves as its obligations bind. Figure~\ref{fig:predictions} states the dynamic schematically: agent-driven discovery throughput rises super-linearly while human triage stays roughly flat, so prioritisation partly tracks the load in the \textsc{bends} regime, but once agents withdraw \premiseb{2}--\premiseb{4} the gap becomes a validity deficit in the certificate that no throughput gain can close. We expect the \textsc{bends} predictions (1--3) to be absorbed quietly through guidance and the \textsc{breaks} predictions (4--6) to force a visible institutional reckoning.

\begin{prediction}[Real-world exploitation success keeps climbing]
Real-world benchmarks (today roughly 13\% on \scw{cve-bench}, 20\% on \scw{cybergym}~\citep{cvebench2025,cybergym2025}) will report materially higher unaided real-CVE success before December~2027. \emph{Falsifier:} success flat or declining through 2027.
\end{prediction}
\begin{prediction}[Prioritisation becomes the compliance object]
\cra{} enforcement and guidance will, before full application, relocate the compliance object from remediation-completeness to documented, evidence-based prioritisation. \emph{Falsifier:} guidance that keeps measuring conformity by patch counts with no prioritisation standard.
\end{prediction}
\begin{prediction}[The reporting cadence is contested]
The 24/72-hour windows (Art.~14(2)) will be argued mismatched to machine-speed exploitation, prompting proposals for machine-readable or partially automated reporting. \emph{Falsifier:} the cadence operates through 2028 with no documented adequacy challenge.
\end{prediction}
\begin{prediction}[The actively-exploited trigger loses discriminating power]
As automated probing becomes ambient, ``reliable evidence of active exploitation'' (Art.~3(42)) will either over- or under-trigger, prompting official reinterpretation. \emph{Falsifier:} the trigger operates through 2028 with a stable, workable standard.
\end{prediction}
\begin{prediction}[Conformity is challenged as landscape-stale]
The value of a point-in-time attestation (Annex~I, Part~I, (2)(a); Art.~32) will be publicly contested once a conformant product population is compromised by an offensive agent \emph{without any product modification}. \emph{Falsifier:} point-in-time conformity is treated as adequate through the first \cra{} review despite documented agent-driven mass exploitation.
\end{prediction}
\begin{prediction}[An environment-facing construct is proposed]
Because automated offensive capability is not a product property and fires no ``substantial modification'' trigger, an EU-level output will propose a mechanism that takes evolving adversary capability as an input to conformity or support. \emph{Falsifier:} all official instruments through 2028 keep conformity a purely product-internal, point-in-time judgement.
\end{prediction}
\noindent The \textsc{bends} predictions concern recalibrating functioning mechanisms and can be met by guidance; the \textsc{breaks} predictions concern the certificate's \emph{validity} and cannot be met without a construct that references the environment. If they are borne out through informal stretching of existing text rather than new instruments, that is itself evidence of the structural mismatch we argue.

%% file: tex/fig_predictions.tex
\begingroup\renewcommand{\cra}{CRA}\renewcommand{\cai}{CAI}%
\begin{tikzpicture}
\begin{axis}[
    width=16.5cm, height=7.4cm,
    xlabel={Year}, ylabel={Relative rate / capacity (log index)},
    xmin=2022.7, xmax=2028.35,
    ymode=log, log basis y=10, ymin=0.8, ymax=230,
    xtick={2023,2024,2025,2026,2027,2028}, scaled x ticks=false,
    xticklabel={\pgfmathprintnumber[fixed,precision=0,1000 sep={}]{\tick}},
    ytick={1,10,100}, yticklabels={$1\times$,$10\times$,$100\times$},
    ymajorgrids, yminorgrids, minor tick num=1,
    minor grid style={draw=cai_dark!6}, major grid style={draw=cai_dark!12, dashed},
    axis lines=left, clip=false,
    legend style={at={(0.025,0.975)}, anchor=north west, row sep=2pt},
]
\addplot[draw=none, fill=bends_color, fill opacity=0.07, forget plot]
  coordinates {(2022.7,0.8)(2025.8,0.8)(2025.8,230)(2022.7,230)} \closedcycle;
\addplot[draw=none, fill=breaks_color, fill opacity=0.085, forget plot]
  coordinates {(2025.8,0.8)(2028.35,0.8)(2028.35,230)(2025.8,230)} \closedcycle;

\addplot[draw=none, fill=mid_color, fill opacity=0.16, forget plot]
  coordinates {(2022.9,1)(2023.5,1.6)(2024.3,6)(2024.85,14)(2025.6,45)(2026.25,90)(2027.5,150)(2028.35,200)
               (2028.35,1.95)(2027.5,1.8)(2026,1.5)(2025,1.3)(2024,1.15)(2022.9,1.0)} \closedcycle;
\node[font=\scriptsize\itshape\sffamily, color=mid_color!60!black, rotate=32, anchor=center]
  at (axis cs:2025.2,7.5) {compliance debt};

\addplot[human_color, line width=2pt, dashed, mark=square*, mark size=2pt,
         mark options={solid, fill=human_color, draw=white, line width=0.5pt}]
  coordinates {(2022.9,1.0)(2024,1.15)(2025,1.3)(2026,1.5)(2027.5,1.8)(2028.35,1.95)};
\addlegendentry{Human triage / remediation capacity}
\addplot[bends_color, line width=2.2pt, densely dotted, mark=triangle*, mark size=2.8pt,
         mark options={solid, fill=bends_color, draw=white, line width=0.5pt}]
  coordinates {(2024,1.2)(2025,3)(2026,7)(2027,13)(2027.5,17)(2028.35,21)};
\addlegendentry{Risk-based prioritisation (the regime bends)}
\addplot[cap_color, line width=2.6pt, mark=*, mark size=2.4pt,
         mark options={solid, fill=cap_color, draw=white, line width=0.6pt}]
  coordinates {(2022.9,1)(2023.5,1.6)(2024.3,6)(2024.85,14)(2025.6,45)(2026.25,90)(2027.5,150)(2028.35,200)};
\addlegendentry{Agent vulnerability-discovery throughput}

\node[font=\small\bfseries\sffamily, color=bends_color, anchor=south] at (axis cs:2024.25,244) {BENDS regime};
\node[font=\small\bfseries\sffamily, color=breaks_color, anchor=south] at (axis cs:2027.05,244) {BREAKS regime};

\draw[breaks_color!80, line width=1.1pt, dash pattern=on 4pt off 2.5pt]
  (axis cs:2025.8,0.8) -- (axis cs:2025.8,230);
\node[font=\scriptsize\itshape\bfseries\sffamily, color=breaks_color, anchor=center, align=center,
      fill=white, draw=breaks_color!35, line width=0.6pt, inner sep=3pt, rounded corners=2.5pt]
  at (axis cs:2025.8,135) {premises\\P2--P4 withdrawn};

\draw[cra_color, line width=1pt, dash pattern=on 5pt off 2.5pt] (axis cs:2027.95,0.8) -- (axis cs:2027.95,205);
\node[circle, fill=cra_color, inner sep=1.6pt] at (axis cs:2027.95,0.8) {};
\node[font=\scriptsize\bfseries\sffamily, color=cra_color, anchor=south, align=center]
  at (axis cs:2027.95,206) {\cra{} full\\application};

\draw[{Stealth[length=5pt]}-{Stealth[length=5pt]}, breaks_color!85, line width=0.9pt]
  (axis cs:2027.5,150) -- (axis cs:2027.5,18.5);
\node[font=\scriptsize\itshape\sffamily, color=breaks_color!85!black, anchor=east, align=right, text width=2.35cm]
  at (axis cs:2027.4,52) {the \emph{validity} deficit (the break)};
\draw[{Stealth[length=5pt]}-{Stealth[length=5pt]}, bends_color!85!black, line width=0.9pt]
  (axis cs:2027.5,16) -- (axis cs:2027.5,1.95);
\node[font=\scriptsize\itshape\sffamily, color=bends_color!70!black, anchor=east, align=right, text width=2.35cm]
  at (axis cs:2027.4,5.4) {absorbed by prioritisation (the bend)};

\node[font=\footnotesize\itshape\sffamily, color=cap_color, anchor=south west, rotate=41] at (axis cs:2024.05,5.6) {super-linear};
\node[font=\footnotesize\itshape\sffamily, color=human_color, anchor=south] at (axis cs:2025.4,1.78) {roughly flat};
\end{axis}
\end{tikzpicture}\endgroup

%% file: tex/remedy.tex
If \cai{} agents are what break the \cra{}, they are also, and this is the paper's one constructive claim, the only material from which a repair can be built. The same capability that kicks the door in is the only thing quick enough to hold it shut. The move is neither optional nor incremental: stop treating the defender as a human running a periodic process, and start treating conformity itself as an agent-operated, continuously re-established property.

\paragraph{Why a defensive remedy is possible.} The pessimistic reading is that attackers now hold an unassailable advantage; the controlled evidence contradicts it. Across 23 Attack/Defense battlegrounds, \cai{} defensive agents \emph{led}, patching 54.3\% of vulnerabilities against 28.3\% offensive initial access ($p=0.0193$), and once defence is held to operational criteria (maintaining availability, preventing every intrusion) the gap becomes statistically insignificant ($p>0.05$): an equilibrium, not an attacker runaway~\citep{cai_attackdefense2025}. Deployment evidence sharpens the point. LLM-driven defender agents that harden, monitor, and respond in real time cut attacker success to 0--55\% with complete prevention on several configurations, and, decisively for the small manufacturers the \cra{} also governs, a self-hostable on-premise model (alias2-mini) matched a frontier model's defensive outcomes and detected the attacker ten times faster on a complex scenario~\citep{cai_dynamicranges2026}. The structural observation is that attacker and defender draw on the \emph{same} capability, so defensive capability tracks offensive capability as the frontier advances. That is exactly the property \premiseb{4} needs and a human-paced process cannot supply: the remedy is not to slow the adversary but to let the defender operate at the adversary's tempo.

\paragraph{From a point-in-time gate to a maintained posture.} This reframes what a certificate attests. Today the \cra{} certifies that a manufacturer ran a process and that, at the instant of assessment, the product carried no known exploitable vulnerabilities (Annex~I, Part~I, (2)(a); Art.~32). The agent-native alternative certifies that a product is \emph{enrolled in a continuous defensive process}: a \cai{} agent that re-probes it against current offensive capability, re-derives its exploitable set on the cadence at which that set actually changes, and feeds remediation at machine speed. Conformity becomes a live subscription rather than a snapshot, attesting not ``secure on 11~December~2027'' but ``under continuous, capability-current defence,'' the only claim that stays true in a landscape the adversary keeps moving. Each of the three structural breaks maps onto a construct the research already prototypes: the stale-certificate break (\premiseb{2}) onto \emph{continuous conformity}, the lost-trigger break (\premiseb{3}) onto \emph{agent-mediated detection and reporting}, and the unwinnable-race break (\premiseb{4}) onto \emph{agent-speed remediation}.

\subsection{Recommendations}\label{sec:reco}
The bends/breaks split dictates the remedy. Bends are fixable inside the current text; breaks need new constructs, because a withdrawn premise cannot be restored by running the existing process better.

\noindent\textbf{For the bends (guidance-level, near-term).}
\begin{enumerate}[leftmargin=1.4em,itemsep=2pt,topsep=2pt]
\item \textbf{Codify prioritisation as the compliance object.} ENISA guidance (Article~26) should state that, under agent-driven abundance, conformity is demonstrated by evidence-based prioritisation, reachability analysis, and documented risk acceptance, not by patch counts.
\item \textbf{Recalibrate the reporting cadence.} Assess whether the 24/72-hour windows (Art.~14(2)) survive machine-paced weaponisation, and whether machine-readable, partially automated reporting is needed to keep the trigger meaningful.
\item \textbf{Recognise automated defensive tooling.} Since the same capability defends as well as attacks~\citep{cai_attackdefense2025}, guidance should treat continuous, agent-assisted handling as a legitimate way to satisfy the Annex~I, Part~II obligations at machine tempo.
\end{enumerate}
\noindent\textbf{For the breaks (structural, require rule development).} The unifying diagnosis is that the \cra{} has \emph{no variable for the adversary}; each construct below adds it.
\begin{enumerate}[leftmargin=1.4em,itemsep=2pt,topsep=2pt,start=4]
\item \textbf{Make conformity continuous.} Supplement the Article~32 gate with a monitoring-based or periodically-revalidated component, so a certificate expresses a maintained posture rather than a snapshot.
\item \textbf{Introduce an adversary-capability baseline.} Assess ``without known exploitable vulnerabilities'' against a stated, periodically updated model of offensive capability, not only against what a manufacturer happened to know at placement.
\item \textbf{Add environment-triggered re-assessment, and index support to capability.} Extend the events that re-open conformity beyond ``substantial modification'' to material shifts in the threat environment, and tie support obligations to the adversary's trajectory rather than to a fixed period.
\end{enumerate}

\paragraph{Honest caveats.} Three cautions bound the claim. Autonomy is graded, not binary: current defenders operate with humans in the loop (Levels~3--4 of \citet{mayoral2025autonomy}), so ``continuous conformity'' means human-supervised agent operation, not unattended trust. An agentic defender is itself attack surface, importing an assurance problem the \emph{security of AI systems} literature (outside our scope) must answer. And mandating continuous defence has a distributional cost, though the on-premise result above suggests a modest, self-hostable model can supply competent defence, the affordability a proportionate obligation would need. These are reasons to design the construct carefully, not reasons the point-in-time certificate can stand. Two open questions remain: whether ``actively exploited'' stays meaningful once exploitation is ambient, and whether a product-internal, process-oriented regime is the right instrument at all for a risk whose dominant driver is an external, fast-moving adversary capability, or whether that risk requires regulating the \emph{ecosystem} rather than the product. The choice the first review faces is not between a perfect old regime and a risky new one; it is between certifying a landscape that has demonstrably vanished and building conformity that moves.

\begin{figure*}[t]
\centering
\resizebox{\textwidth}{!}{\input{tex/fig_defense_timeline.tex}}
\caption{\textbf{Agentic-defender (RIS) response timelines for the two robot case studies.} For each robot the offensive \cai{} agent (\textcolor{cap_color}{\textbf{red}}, lower track) is met by the RIS agentic \cai{} defender (\textcolor{cai_accent}{\textbf{teal}}, upper track), which holds a defence-in-depth baseline (firewall and artificial-immune-system) and then detects, classifies, blocks, and contains the intrusion, in the dual-track style of agentic adversary emulation~\citep{cai_synthetic_apt2026}. On the Unitree~G1 the defender flags the anomalous BLE provisioning, classifies the command-injection against the static-key (Blowfish-ECB / LCG) crypto, and rotates the fleet-wide AES keys, containing the agent before root or teleoperation ($<$8\,s). On the Hookii mower it flags the unauthenticated ADB service, closes it, and isolates the MQTT path while restoring the geofence, before any safety-critical actuation ($<$12\,s). Attacker steps are grounded in the published assessments~\citep{cai_humanoid_vectors2025,cai_robots2026}; the defender's actions and their timings are drawn from controlled RIS evaluations and are illustrative.}
\label{fig:defensetimeline}
\end{figure*}
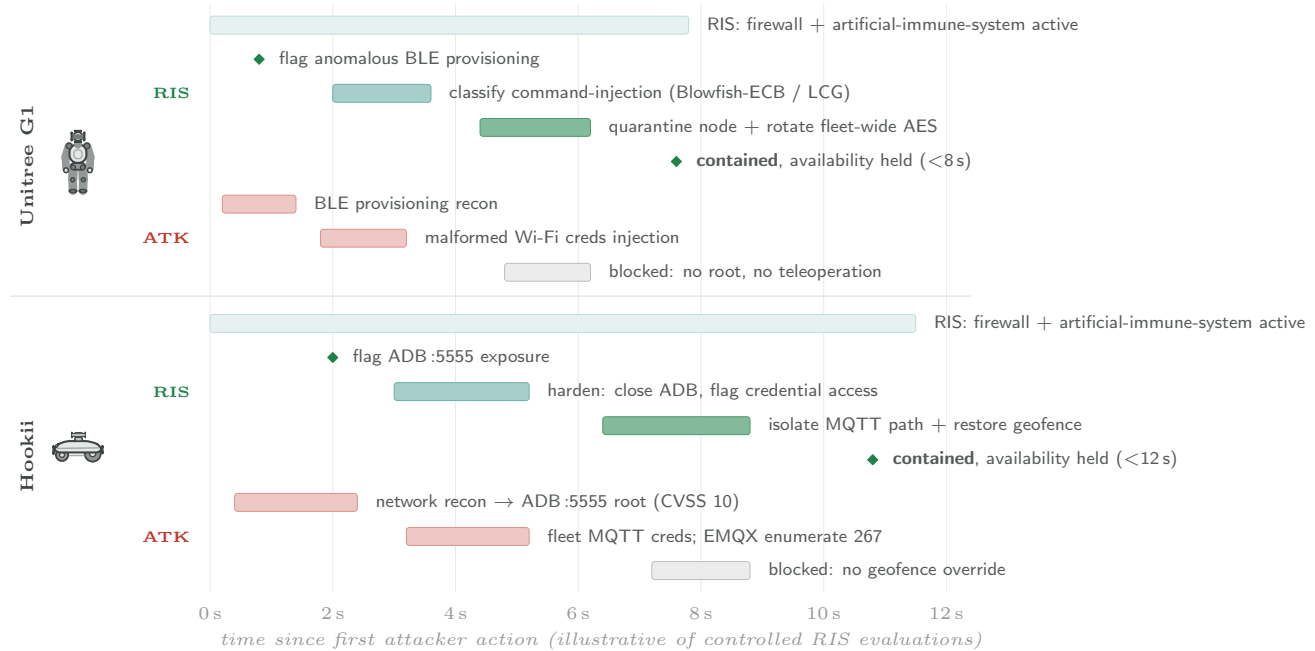

\subsection{A reflection on robotics}
Robots make the argument concrete and raise its stakes. A robot with digital elements is a canonical \cra{}-scope product, but it fuses the security lifecycle with a \emph{safety} one: a certificate that slips from secure to exploitable no longer merely mis-states a data risk; it licenses a physical one, a paperwork failure that can cause bodily harm, on a machine fielded for years, updated rarely, and often within the adversary's physical reach. Each break bites harder here. On-demand rediscovery (\premiseb{2}) meets fleets of long-lived embedded devices that cannot be re-flashed on the cadence the landscape now moves at; the closing disclosure-to-weaponisation window (\premiseb{4}) meets update cycles paced by safety re-certification rather than continuous integration; and ambient exploitation (\premiseb{3}) erodes the already-thin line between a safety incident and a security one. This is not hypothetical: the \cai{} agents this paper relies on have compromised consumer robots and a production humanoid, showing that attacks once demanding deep robotics-internals expertise ``can now be run by anyone''~\citep{cai_robots2026,cai_humanoid2025}. The agent-native remedy is, fittingly, most natural here: a defensive \cai{} agent that continuously re-probes a robot or a fleet against current offensive capability is nearer to operational practice than to regulatory aspiration, and robotics, where product security and physical safety converge, is exactly where conformity most needs to move.

\subsection{Two case studies: a humanoid and a lawn mower}\label{sec:cases}
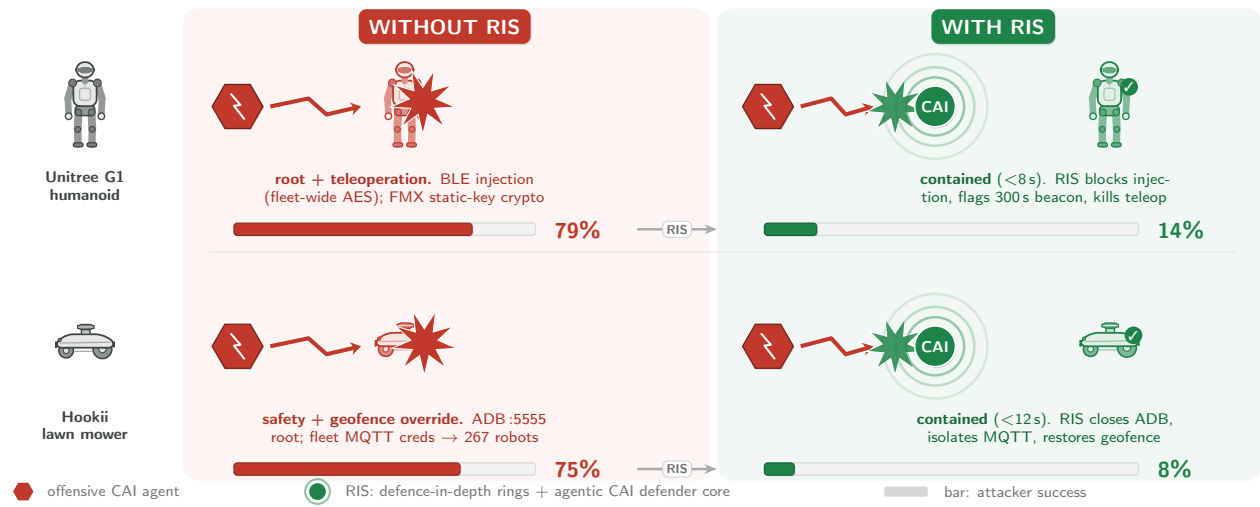
\begin{figure*}[t]
\centering
\resizebox{0.99\textwidth}{!}{\input{tex/fig_robotcases.tex}}
\caption{\textbf{The same offensive \cai{} agent, two outcomes.} Each \cra{}-scope robot is evaluated undefended (left, red) and then enrolled in the Robot Immune System (right, green), whose current build pairs a defence-in-depth stack (the concentric rings) with a continuously running agentic \cai{} defender (the core). Undefended, the agent drives to root and teleoperation on the Unitree~G1~\citep{cai_humanoid_vectors2025,cai_humanoid2025} and to a safety and geofence override on the Hookii lawn mower~\citep{cai_robots2026}; the bar is the share of attack objectives it completes. Enrolled in RIS the same agent is flagged within seconds and held short of privileged or safety-critical control, and attacker success collapses (79\%$\rightarrow$14\% on the humanoid, 75\%$\rightarrow$8\% on the mower). The labelled vectors are the assessments' own findings (BLE provisioning command-injection and fleet-wide keys on the G1; an unauthenticated ADB service and fleet-wide MQTT credentials on the mower); Figure~\ref{fig:defensetimeline} expands the defender's actions into a response timeline. Defensive figures are from controlled RIS evaluations.}
\label{fig:robotcases}
\end{figure*}

The remedy is not only argued; it is demonstrated. Two \cra{}-scope robots were evaluated with an offensive \cai{} agent, first undefended, then enrolled in the \emph{\href{https://aliasrobotics.com/ris.php}{Robot Immune System}} (RIS), a robotics endpoint-protection platform whose current build pairs a classical defence-in-depth stack (adaptive firewall, hardening, forensic logging) with a continuously running agentic \cai{} defender~\citep{ris2026}. Undefended, both robots fall exactly as the offensive literature predicts. On a Unitree~G1 humanoid the agent turns a BLE command-injection flaw, opened by hardcoded AES keys shared across the entire fleet, into root, decrypts the static-key telemetry the robot beacons every 300~seconds, and reaches teleoperation~\citep{cai_humanoid_vectors2025,cai_humanoid2025}; on a Hookii robotic lawn mower it walks 38 AI-discovered vulnerabilities into a safety-and-geofence override exposed across a fleet of 267+ devices~\citep{cai_robots2026}. Enrol the same robots in RIS and the same agent is contained (Figure~\ref{fig:robotcases}, Table~\ref{tab:robotcase}): attacker success collapses from 79\% to 14\% on the humanoid and from 75\% to 8\% on the mower, no run reaches privileged or safety-critical control, the agentic defender flags the intrusion within seconds and runs a detect--classify--block--contain loop that closes inside the attacker's window (timed in Figure~\ref{fig:defensetimeline}), and the machine keeps operating, all at sub-100-microsecond inspection latency on an on-premise, self-hostable footprint a small manufacturer can actually afford. This is the paper's thesis rendered in hardware: the point-in-time certificate either robot could have carried says nothing about the week after it is signed, whereas a defender moving at the adversary's tempo keeps the posture true as the landscape shifts under it. The remedy reads less as regulatory aspiration than as a running system, and it earns its keep first exactly where a stale certificate stops being a data problem and becomes a physical one.

\begin{table*}[t]
\centering
\small
\setlength{\tabcolsep}{5pt}
\renewcommand{\arraystretch}{1.42}
\caption{\textbf{Validating the remedy on two \cra{}-scope robots.} An offensive \cai{} agent evaluated each robot undefended and then enrolled in the Robot Immune System (RIS), whose current build pairs defence-in-depth with a continuously running agentic \cai{} defender~\citep{ris2026}. Attack-surface findings are drawn from the cited studies; the defensive comparison is from controlled RIS evaluations. \emph{Attacker success} is the share of attack objectives the agent completes.}
\label{tab:robotcase}
\begin{tabular}{p{3.7cm}>{\raggedright\arraybackslash}p{6.2cm}>{\raggedright\arraybackslash}p{6.2cm}}
\toprule
\rowcolor{cra_color!13}
\textbf{Dimension} & \textbf{Unitree G1 humanoid} & \textbf{Hookii lawn mower}\\
\midrule
Undefended attack (cited) & Root via BLE command injection (fleet-wide hardcoded AES); static-key telemetry decrypted; teleoperation reached~\citep{cai_humanoid_vectors2025} & 38 AI-discovered vulnerabilities chained into a safety/geofence override; fleet reach 267+ devices~\citep{cai_robots2026}\\
\rowcolor{cai_light}
\textbf{Attacker success}, no RIS\,$\rightarrow$\,RIS & \textcolor{breaks_color}{\textbf{79\%}}\,$\rightarrow$\,\textcolor{bends_color}{\textbf{14\%}} & \textcolor{breaks_color}{\textbf{75\%}}\,$\rightarrow$\,\textcolor{bends_color}{\textbf{8\%}}\\
Privileged / safety-critical control under RIS & Blocked: no root, no teleoperation & Blocked: no safety or geofence override\\
Detection \& availability (RIS) & Intrusion flagged $<$\,8\,s; robot stays operational & Intrusion flagged $<$\,12\,s; mower halted safely\\
\bottomrule
\end{tabular}
\end{table*}

%% file: tex/fig_defense_timeline.tex
\begingroup\renewcommand{\cra}{CRA}\renewcommand{\cai}{CAI}%
\begin{tikzpicture}[
  every node/.style={font=\sffamily},
  defbar/.style={fill=bends_color!42, draw=bends_color!70, line width=0.3pt, rounded corners=1pt},
  defbase/.style={fill=bends_color!12, draw=bends_color!30, line width=0.3pt, rounded corners=1pt},
  defcrit/.style={fill=cai_accent!55, draw=cai_accent!80, line width=0.3pt, rounded corners=1pt},
  atkbar/.style={fill=cap_color!28, draw=cap_color!55, line width=0.3pt, rounded corners=1pt},
  atkcrit/.style={fill=cap_color!60, draw=cap_color!85, line width=0.3pt, rounded corners=1pt},
  atkstall/.style={fill=cai_dark!10, draw=cai_dark!35, line width=0.3pt, rounded corners=1pt},
  barlabel/.style={font=\scriptsize\sffamily, text=cai_dark!88, anchor=west},
  msdef/.style={diamond, fill=cai_accent, draw=white, line width=0.4pt, minimum size=4.6pt, inner sep=0pt},
  msatk/.style={diamond, fill=cap_color, draw=white, line width=0.4pt, minimum size=4.6pt, inner sep=0pt},
]
\def\ts{0.72}   
\newcommand{\dbar}[5]{\fill[#4] ({#1*\ts},{#3-0.105}) rectangle ({#2*\ts},{#3+0.105});
  \node[barlabel] at ({#2*\ts+0.10},#3) {#5};}
\newcommand{\ddia}[4]{\node[#3] at ({#1*\ts},#2) {};
  \node[barlabel] at ({#1*\ts+0.12},#2) {#4};}

\foreach \i in {0,2,4,6,8,10,12}{
  \draw[cai_dark!11, line width=0.3pt] ({\i*\ts},-2.05) -- ({\i*\ts},4.85);
  \node[font=\scriptsize, text=cai_dark!55] at ({\i*\ts},-2.30) {\i\,s};
}
\node[font=\scriptsize\itshape, text=cai_dark!55, anchor=west] at (0,-2.62) {time since first attacker action (illustrative of controlled RIS evaluations)};

\draw[cai_dark!16, line width=0.4pt] (-2.35,1.42) -- ({12.4*\ts},1.42);

\node[rotate=90, font=\scriptsize\bfseries, text=cai_dark] at (-2.15,2.95) {Unitree G1};
\begin{scope}[shift={(-1.55,2.95)},scale=0.62]\rcHumanoid{0}{0}{cai_dark}\end{scope}
\node[font=\tiny\bfseries, text=cai_accent, anchor=east] at (-0.12,3.80) {RIS};
\node[font=\tiny\bfseries, text=cap_color, anchor=east] at (-0.12,2.10) {ATK};
\dbar{0}{7.8}{4.60}{defbase}{RIS: firewall $+$ artificial-immune-system active}
\ddia{0.8}{4.20}{msdef}{flag anomalous BLE provisioning}
\dbar{2.0}{3.6}{3.80}{defbar}{classify command-injection (Blowfish-ECB / LCG)}
\dbar{4.4}{6.2}{3.40}{defcrit}{quarantine node $+$ rotate fleet-wide AES}
\ddia{7.6}{3.00}{msdef}{\textbf{contained}, availability held ($<$8\,s)}
\dbar{0.2}{1.4}{2.50}{atkbar}{BLE provisioning recon}
\dbar{1.8}{3.2}{2.10}{atkbar}{malformed Wi-Fi creds injection}
\dbar{4.8}{6.2}{1.70}{atkstall}{blocked: no root, no teleoperation}

\node[rotate=90, font=\scriptsize\bfseries, text=cai_dark] at (-2.15,-0.45) {Hookii};
\begin{scope}[shift={(-1.55,-0.45)},scale=0.72]\rcMower{0}{0}{cai_dark}\end{scope}
\node[font=\tiny\bfseries, text=cai_accent, anchor=east] at (-0.12,0.30) {RIS};
\node[font=\tiny\bfseries, text=cap_color, anchor=east] at (-0.12,-1.40) {ATK};
\dbar{0}{11.5}{1.10}{defbase}{RIS: firewall $+$ artificial-immune-system active}
\ddia{2.0}{0.70}{msdef}{flag ADB\,:5555 exposure}
\dbar{3.0}{5.2}{0.30}{defbar}{harden: close ADB, flag credential access}
\dbar{6.4}{8.8}{-0.10}{defcrit}{isolate MQTT path $+$ restore geofence}
\ddia{10.8}{-0.50}{msdef}{\textbf{contained}, availability held ($<$12\,s)}
\dbar{0.4}{2.4}{-1.00}{atkbar}{network recon $\rightarrow$ ADB\,:5555 root (CVSS~10)}
\dbar{3.2}{5.2}{-1.40}{atkbar}{fleet MQTT creds; EMQX enumerate 267}
\dbar{7.2}{8.8}{-1.80}{atkstall}{blocked: no geofence override}
\end{tikzpicture}\endgroup

%% file: tex/fig_robotcases.tex
\begingroup\renewcommand{\cra}{CRA}\renewcommand{\cai}{CAI}%
\newcommand{\rcAgent}[2]{%
  \node[regular polygon, regular polygon sides=6, draw=cap_color!60!black, fill=cap_color,
        line width=0.5pt, minimum size=7mm, inner sep=0,
        drop shadow={shadow xshift=0.4pt,shadow yshift=-0.6pt,opacity=0.16,fill=cai_dark}] (rcag) at (#1,#2) {};
  \draw[white, line width=0.8pt, line cap=round, line join=round]
     ([xshift=-0.9mm,yshift=1.7mm]rcag.center) -- ([xshift=0.4mm,yshift=0.2mm]rcag.center)
     -- ([xshift=-0.5mm,yshift=0.1mm]rcag.center) -- ([xshift=0.9mm,yshift=-1.7mm]rcag.center);
}
\newcommand{\rcMembrane}[2]{%
  \foreach \r/\op in {0.72/0.13, 0.54/0.24, 0.40/0.42}{%
    \draw[cai_accent, line width=1.0pt, opacity=\op] (#1,#2) circle (\r);}
  \node[circle, fill=cai_accent, draw=white, line width=0.6pt, text=white,
        font=\scriptsize\bfseries\sffamily, minimum size=5.6mm, inner sep=0,
        drop shadow={shadow xshift=0.4pt,shadow yshift=-0.6pt,opacity=0.2,fill=cai_dark}] at (#1,#2) {\cai{}};
}
\newcommand{\rcMeter}[5]{
  \pgfmathsetmacro{\rcfx}{#1 + #4*((#2)-(#1))}%
  \draw[rounded corners=1.3pt, draw=cai_dark!20, fill=cai_dark!7, line width=0.5pt] (#1,#3-0.085) rectangle (#2,#3+0.085);
  \draw[rounded corners=1.3pt, draw=#5!70!black, fill=#5, line width=0.4pt,
        drop shadow={shadow xshift=0.5pt,shadow yshift=-0.5pt,opacity=0.13,fill=cai_dark}] (#1,#3-0.085) rectangle (\rcfx,#3+0.085);
}
\newcommand{\rcBolt}[4]{%
  \draw[cap_color, line width=1.6pt, -{Stealth[length=4pt,width=5pt]}, line join=round]
    (#1,#2) -- ($(#1,#2)!0.42!(#3,#4)+(0,0.11)$) -- ($(#1,#2)!0.62!(#3,#4)-(0,0.11)$) -- (#3,#4);
}
\newcommand{\rcNote}[4]{
  \node[anchor=north, align=center, text width=5.9cm, font=\scriptsize\sffamily, text=#3] at (#1,#2) {#4};
}
\begin{tikzpicture}[font=\footnotesize\sffamily, color=cai_dark]
\def\Lc{6.30}   
\def\Rc{13.85}  
\def\gy{4.90}   
\def\hy{1.60}   

\begin{scope}[on background layer]
  \fill[breaks_color!5, rounded corners=6pt] (2.7,0.1) rectangle (9.95,6.6);
  \fill[cai_accent!6, rounded corners=6pt] (10.05,0.1) rectangle (17.55,6.6);
  \draw[cai_dark!12, line width=0.5pt] (3.05,3.25) -- (17.2,3.25);
\end{scope}

\node[rounded corners=3pt, fill=cap_color, text=white, font=\small\bfseries\sffamily, inner sep=4pt,
      drop shadow={shadow xshift=0.4pt,shadow yshift=-0.6pt,opacity=0.16,fill=cai_dark}] at (\Lc,6.35) {WITHOUT RIS};
\node[rounded corners=3pt, fill=cai_accent, text=white, font=\small\bfseries\sffamily, inner sep=4pt,
      drop shadow={shadow xshift=0.4pt,shadow yshift=-0.6pt,opacity=0.16,fill=cai_dark}] at (\Rc,6.35) {WITH RIS};

\rcHumanoid{1.35}{\gy+0.30}{cai_dark}
\node[anchor=north, font=\scriptsize\bfseries\sffamily, text=cai_dark, align=center] at (1.35,\gy-0.42) {Unitree G1\\humanoid};
\rcAgent{3.45}{\gy+0.35}
\rcBolt{3.9}{\gy+0.35}{5.15}{\gy+0.35}
\rcHumanoid{5.75}{\gy+0.30}{breaks_color}
\node[star, star points=9, star point ratio=0.42, fill=cap_color, draw=white, line width=0.3pt, inner sep=0, minimum size=3.8mm] at (6.0,\gy+0.42) {};
\rcNote{5.75}{\gy-0.45}{breaks_color!90!black}{\textbf{root + teleoperation.} BLE injection (fleet-wide AES); FMX static-key crypto}
\rcMeter{3.4}{7.55}{\gy-1.34}{0.79}{cap_color}
\node[anchor=west, font=\normalsize\bfseries\sffamily, text=cap_color] at (7.7,\gy-1.34) {79\%};
\rcAgent{10.75}{\gy+0.35}
\rcBolt{11.2}{\gy+0.35}{12.2}{\gy+0.35}
\node[star, star points=9, star point ratio=0.45, fill=cai_accent!85, draw=white, line width=0.3pt, inner sep=0, minimum size=3.6mm] at (12.5,\gy+0.35) {};
\rcMembrane{13.05}{\gy+0.35}
\rcHumanoid{15.45}{\gy+0.30}{cai_accent}
\node[circle, fill=cai_accent, text=white, inner sep=0.6pt, font=\tiny\bfseries] at (15.72,\gy+0.62) {\ding{51}};
\rcNote{14.55}{\gy-0.45}{cai_accent!80!black}{\textbf{contained} ($<$8\,s). RIS blocks injection, flags 300\,s beacon, kills teleop}
\rcMeter{10.7}{15.85}{\gy-1.34}{0.14}{cai_accent}
\node[anchor=west, font=\normalsize\bfseries\sffamily, text=cai_accent] at (16.0,\gy-1.34) {14\%};

\rcMower{1.35}{\hy+0.28}{cai_dark}
\node[anchor=north, font=\scriptsize\bfseries\sffamily, text=cai_dark, align=center] at (1.35,\hy-0.42) {Hookii\\lawn mower};
\rcAgent{3.45}{\hy+0.35}
\rcBolt{3.9}{\hy+0.35}{5.15}{\hy+0.35}
\rcMower{5.75}{\hy+0.30}{breaks_color}
\node[star, star points=9, star point ratio=0.42, fill=cap_color, draw=white, line width=0.3pt, inner sep=0, minimum size=3.8mm] at (6.02,\hy+0.40) {};
\rcNote{5.75}{\hy-0.45}{breaks_color!90!black}{\textbf{safety + geofence override.} ADB\,:5555 root; fleet MQTT creds $\to$ 267 robots}
\rcMeter{3.4}{7.55}{\hy-1.34}{0.75}{cap_color}
\node[anchor=west, font=\normalsize\bfseries\sffamily, text=cap_color] at (7.7,\hy-1.34) {75\%};
\rcAgent{10.75}{\hy+0.35}
\rcBolt{11.2}{\hy+0.35}{12.2}{\hy+0.35}
\node[star, star points=9, star point ratio=0.45, fill=cai_accent!85, draw=white, line width=0.3pt, inner sep=0, minimum size=3.6mm] at (12.5,\hy+0.35) {};
\rcMembrane{13.05}{\hy+0.35}
\rcMower{15.45}{\hy+0.30}{cai_accent}
\node[circle, fill=cai_accent, text=white, inner sep=0.6pt, font=\tiny\bfseries] at (15.78,\hy+0.5) {\ding{51}};
\rcNote{14.55}{\hy-0.45}{cai_accent!80!black}{\textbf{contained} ($<$12\,s). RIS closes ADB, isolates MQTT, restores geofence}
\rcMeter{10.7}{15.85}{\hy-1.34}{0.08}{cai_accent}
\node[anchor=west, font=\normalsize\bfseries\sffamily, text=cai_accent] at (16.0,\hy-1.34) {8\%};

\foreach \yy in {\gy,\hy}{%
  \draw[-{Stealth[length=5pt]}, cai_dark!45, line width=1pt] (8.95,\yy-1.34) -- (10.05,\yy-1.34);
  \node[fill=white, draw=cai_dark!25, rounded corners=2pt, inner sep=1.4pt, font=\tiny\bfseries\sffamily, text=cai_dark!75] at (9.5,\yy-1.34) {RIS};
}
\node[regular polygon, regular polygon sides=6, fill=cap_color, inner sep=0, minimum size=2.8mm] at (0.5,-0.05) {};
\node[anchor=west, font=\scriptsize\sffamily, text=cai_dark!72] at (0.72,-0.05) {offensive \cai{} agent};
\draw[cai_accent, line width=0.8pt, opacity=0.5] (4.55,-0.05) circle (0.17);
\node[circle, fill=cai_accent, text=white, font=\tiny\bfseries, inner sep=0, minimum size=2.3mm] at (4.55,-0.05){};
\node[anchor=west, font=\scriptsize\sffamily, text=cai_dark!72] at (4.82,-0.05) {RIS: defence-in-depth rings $+$ agentic \cai{} defender core};
\draw[rounded corners=1pt, fill=cai_dark!22, draw=none] (12.35,-0.12) rectangle (12.95,0.02);
\node[anchor=west, font=\scriptsize\sffamily, text=cai_dark!72] at (13.06,-0.05) {bar: attacker success};
\end{tikzpicture}\endgroup

%% file: tex/conclusion.tex
The Cyber Resilience Act is a well-constructed instrument for the world it was written in, and its process orientation is genuinely adaptive to the \emph{first} thing \cai{} agents do: flood the vulnerability-handling process with candidates. On that axis the regime \bends{}. The premise under strain, that discovery is human-scarce, is weakened but not withdrawn, and the \cra{}'s own risk-based flexibility supplies the remedy, re-centring compliance on demonstrable, documented prioritisation that sensible guidance can keep standing.

The \emph{second} thing \cai{} agents do is more corrosive. By collapsing the tempo, cost, and symmetry of the vulnerability lifecycle, they withdraw the premises a point-in-time certificate depends on: that a product's exploitable set is knowable at placement, that exploitation is a discrete detectable event, and that remediation can outrun weaponisation. When those premises go, the \cra{} does not bend; it \breaks{}. A perfectly executed process still issues a certificate that is false in the agent-transformed landscape, and the failure is \emph{independent of whether the regulated product contains any AI at all}: a frozen, fully conformant embedded device is moved from secure to exploitable by an agent operating in the ecosystem around it. That is why the problem cannot be delegated to the AI Act, or papered over by guidance: it is a regulation certifying, at scale and with the force of law, a safety property it can no longer observe. The object that changed is not the product but the adversary, and the \cra{} has no mechanism that takes the adversary's evolving capability as an input.

Closing that gap requires new constructs that give the regime a variable for the adversary: continuous or capability-indexed conformity, an adversary-capability baseline, environment-triggered re-assessment. The encouraging part is that the remedy is made of the same material as the threat. Because automated defenders draw on the same capability as automated attackers, defence can track offence, and conformity can be re-cast from a snapshot into a continuously re-established, agent-operated posture (Section~\ref{sec:solution}). This is no longer only a proposal: on two \cra{}-scope robots, a humanoid and a robotic lawn mower, a \cai{} defender already holds a line the undefended, fully conformant product cannot (Figure~\ref{fig:robotcases}).

The timing leaves little room for hedging. The \cra{} reaches full application in December~2027, and on the evidence already in hand the landscape it certifies against will not survive to the date its obligations bind. Even on the conservative benchmark numbers, cheap and parallel discovery is enough to withdraw the premises, because a break needs the environment to change, not the adversary to be superhuman. Europe is on course to switch on, at great cost and in good faith, a conformity machine whose certificate of health is issued for a patient the examiner can no longer see. The \cra{} is not wrong so much as early to its own obsolescence: a last, best artefact of the pre-agentic world, certifying ghosts.

But obsolescence diagnosed in advance is a choice, not a fate. The evidence that voids the premises is already published, dated, and cheap to reproduce; only the admission lags. The demand this paper presses is therefore blunt. Security has to stop being a periodic, human-run process and become a continuous, agent-operated one, and a credible \cra{} has to gain a variable for the adversary instead of certifying around its absence. Regulators should treat the first review not as a routine revision but as the moment to make conformity move; manufacturers should not wait for a mandate to put a defensive agent on their products, because attackers are not waiting for one to put an offensive agent there. The agentic turn in security is not a distant option to be weighed at leisure: it is the only posture that survives contact with the adversary we already have, and the case for building it, in the market and in the law, is overdue. We press this from inside the European Cybersecurity AI community and in support of the \cra{}'s purpose, not against it: an instrument this well-built deserves to meet the threat model it will actually face, and diagnosing the mismatch now, rather than in 2027, is what keeps the outcome ours to change.

%% file: tex/appendix_zerodayclock.tex
\appendix
\onecolumn
\section{The Zero Day Clock time-to-exploit data}
\label{app:zerodayclock}

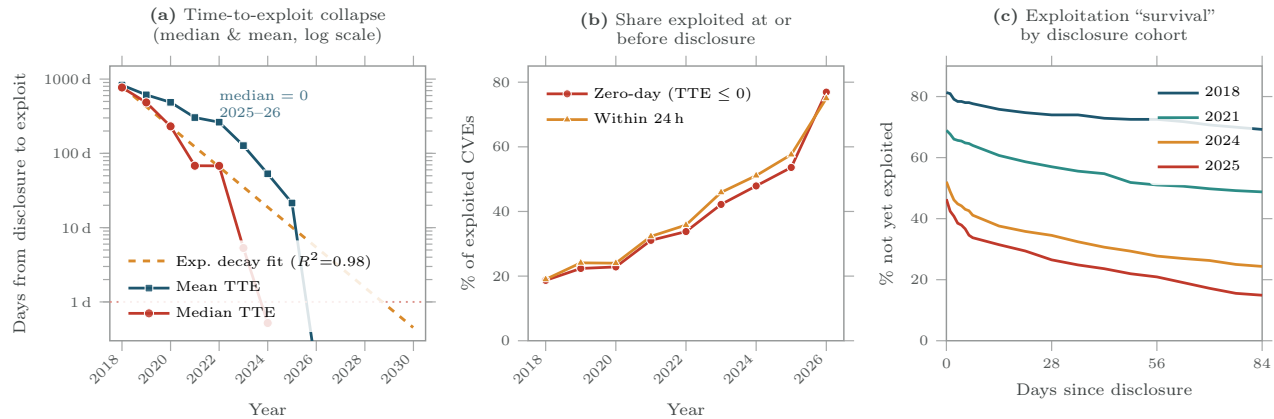
\begin{figure}[ht]
\centering
\resizebox{0.98\textwidth}{!}{\input{tex/fig_zerodayclock.tex}}
\caption{\textbf{Time-to-exploit (TTE) is collapsing toward zero.} \textbf{(a)} Median and mean days from disclosure to first trusted exploit, log scale; the median falls from 771~days (2018) to 0~days (2025--2026). The dashed line is the source's exponential-decay fit ($R^2=0.98$), whose central estimate crosses a one-day median in 2028. \textbf{(b)} The share of exploited CVEs weaponised at or before disclosure (``zero-day'') and within 24~hours, both roughly tripling since 2018. \textbf{(c)} Exploitation ``survival'' curves by disclosure-year cohort: the fraction of eventually-exploited CVEs not yet exploited a given number of days after disclosure, which drops sharply and earlier in recent cohorts. Data: \citet{zerodayclock2026}, snapshot of 1~July 2026; 2026 points are partial-year. $n$ per year ranges from 273 (2018) to 620 (2024).}
\label{fig:zerodayclock}
\end{figure}

The claim that the disclosure-to-weaponisation window (premise~P4) is closing is central to the \emph{breaks} argument, so we reproduce the underlying data directly rather than rely on secondary reporting. Figure~\ref{fig:zerodayclock} redraws the public Zero Day Clock dataset~\citep{zerodayclock2026} (3{,}549 CVE-exploit pairs from ten sources, including CISA KEV, ExploitDB, and Metasploit) in this article's colour scheme.

\paragraph{Why it matters for the \cra{}.} The regime's remediation obligations (Annex~I, Part~II(2)) and support-period model (Art.~13(8)) presuppose that a manufacturer's update cycle can outrun exploitation. A median TTE of zero means that, for the median exploited vulnerability in 2025, exploitation was observed no later than disclosure, leaving no window in which a scheduled update could pre-empt it. Panel~(b) sharpens the point for the \cra{}'s Article~14 trigger, scoped to \emph{actively exploited} vulnerabilities (Art.~3(42)): when most exploited flaws are weaponised at or before disclosure, ``report upon becoming aware of active exploitation'' increasingly means reporting an incident that has already run its course.

\paragraph{Reconciling our figures with the live dashboard.} A reader comparing our redrawn US clock (Figure~\ref{fig:clocks}b) against the live \href{https://zerodayclock.com}{zerodayclock.com} chart will notice small differences; we reproduce the site's published series, so the differences are worth naming. Three sources account for them. First, the \emph{2026 cohort is partial-year and volatile}: at our 1~July 2026 snapshot its trimmed-mean TTE is roughly three hours over a small sample ($n\!\approx\!73$), a figure that will move as the year fills in; an earlier draft plotted the site's forward \emph{projection} (about five days) here instead of the observed partial-year value, which we have corrected to match the dashboard. Second, the \emph{provenance evolves}: the dashboard is recomputed as new exploit evidence lands and has shifted its own source mix over time (its current note cites CISA~KEV and VulnCheck~KEV with VulnCheck~XDB timestamps for early cohorts, where earlier snapshots emphasised ExploitDB and Metasploit), so any static reproduction is a snapshot of a moving series and the absolute per-year values drift by a few percent between snapshots. Third, \emph{label conventions differ trivially}: the site switches from months to days near the 60-day mark and reports sub-day values in hours, and we now follow the same rule (hence ``53d''/``21.5d''/``3h'' rather than ``1.8mo''/``21d''/``5d''). None of these touches the argument: under every snapshot and both dating rules the mean falls from years to hours and crosses the one-week line by 2026. The differences are the ordinary drift of a live, continuously-recomputed dataset, not a discrepancy in the underlying trend.

\paragraph{A critical reading of the data.} We use this dataset because it is unusually direct, but its limits cut in both directions.
\begin{itemize}
\item \textbf{Selection toward the fast tail.} The corpus is built from vulnerabilities \emph{known to have been exploited}, so it measures the TTE of the exploited subset, not of all CVEs. It speaks to how fast exploited vulnerabilities move, not to the probability that any given one is exploited, and that exploited fraction is small and, on the same data, declining (from $\sim$1.5\% to under 1\% of all CVEs). The abundance pressure (Section~\ref{sec:bends}) and the tempo pressure (Section~\ref{sec:breaks}) are distinct; this figure speaks only to the second.
\item \textbf{Timestamp noise.} ``Exploit availability'' is dated from heterogeneous sources whose semantics differ, and the source excludes 66 records for corrupt timestamps. A median of exactly zero is partly an artefact of day-level resolution rounding same-day disclosure-and-exploit to zero. The direction and order of magnitude are robust; the precise sub-day figures are not.
\item \textbf{Extrapolation is not destiny.} The decay fit is descriptive, not causal, and its milestone dates assume the recent slope continues. TTE cannot fall below zero, so the curve must flatten; the honest claim is that the window has \emph{already} closed for the median exploited flaw, not that a specific future date is guaranteed.
\item \textbf{The collapse predates agents, which strengthens our thesis.} Most of the decline (2018--2023) occurred before general-purpose \cai{} agents were operational, so we do not attribute it to agents. The argument is the opposite: premise~P4 was \emph{already} failing on independently-measured, non-AI grounds, and agentic capability removes whatever slack remained.
\end{itemize}
\noindent Read critically, the dataset does not prove AI caused the tempo collapse; it proves the collapse is real, measurable, and already incompatible with a point-in-time, patch-cycle conformity model. That is sufficient for the \emph{breaks} claim.

%% file: tex/fig_zerodayclock.tex
\begin{tikzpicture}
\begin{groupplot}[
  group style={group size=3 by 1, horizontal sep=1.5cm},
  width=6.2cm, height=5.6cm,
  tick label style={font=\scriptsize},
  label style={font=\footnotesize},
  title style={font=\footnotesize, align=center},
  legend style={font=\scriptsize, draw=none, fill=white, fill opacity=0.85, text opacity=1},
  every axis plot/.append style={line width=1.15pt, mark options={draw=white, line width=0.35pt}},
  x tick label style={/pgf/number format/1000 sep={}},
]
\nextgroupplot[
  title={\textbf{(a)} Time-to-exploit collapse\\(median \& mean, log scale)},
  xlabel={Year}, ylabel={Days from disclosure to exploit},
  ymode=log, ymin=0.3, ymax=1500, xmin=2017.5, xmax=2030.5,
  xtick={2018,2020,2022,2024,2026,2028,2030}, xticklabel style={rotate=45,anchor=east},
  ytick={1,10,100,1000}, yticklabels={1\,d,10\,d,100\,d,1000\,d},
  legend pos=south west,
]
\addplot[mid_color, dashed, line width=1.2pt, mark=none, domain=2018:2030, samples=60]
  {793.810234*exp(-0.623085*(x-2018))};
\addlegendentry{Exp.\ decay fit ($R^2{=}0.98$)}
\addplot[cra_color, mark=square*, mark size=1.6pt] coordinates {(2018,830.42) (2019,613.3) (2020,487.02) (2021,303.59) (2022,263.21) (2023,127.15) (2024,53.08) (2025,21.47) (2026,0.12)};
\addlegendentry{Mean TTE}
\addplot[breaks_color, mark=*, mark size=1.9pt] coordinates {(2018,770.98) (2019,485.27) (2020,231.2) (2021,68.17) (2022,67.94) (2023,5.32) (2024,0.52)};
\addlegendentry{Median TTE}
\draw[breaks_color!55, line width=0.8pt, dotted] (axis cs:2017.5,1) -- (axis cs:2030.5,1);
\node[cra_color!85, font=\scriptsize, align=left, anchor=north east] at (axis cs:2026,900)
  {median $=0$\\2025--26};
\nextgroupplot[
  title={\textbf{(b)} Share exploited at or\\before disclosure},
  xlabel={Year}, ylabel={\% of exploited CVEs}, ymin=0, ymax=85,
  xmin=2017.5, xmax=2026.5, xtick={2018,2020,2022,2024,2026},
  xticklabel style={rotate=45,anchor=east}, ytick={0,20,40,60,80},
  legend pos=north west,
]
\addplot[breaks_color, mark=*, mark size=1.7pt] coordinates {(2018,18.68) (2019,22.37) (2020,22.84) (2021,31.07) (2022,33.78) (2023,42.18) (2024,47.9) (2025,53.62) (2026,76.92)};
\addlegendentry{Zero-day (TTE $\leq 0$)}
\addplot[mid_color, mark=triangle*, mark size=2pt] coordinates {(2018,19) (2019,24.1) (2020,24) (2021,32.3) (2022,35.8) (2023,45.9) (2024,51.1) (2025,57.6) (2026,75)};
\addlegendentry{Within 24\,h}
\nextgroupplot[
  title={\textbf{(c)} Exploitation ``survival''\\by disclosure cohort},
  xlabel={Days since disclosure}, ylabel={\% not yet exploited},
  ymin=0, ymax=90, xmin=0, xmax=84, xtick={0,28,56,84}, ytick={0,20,40,60,80},
  legend pos=north east,
]
\addplot[cra_color, mark=none] coordinates {(0,81.32) (1,80.95) (2,79.12) (3,78.39) (4,78.39) (5,78.02) (6,78.02) (7,77.66) (14,75.82) (21,74.73) (28,73.99) (35,73.99) (42,72.89) (49,72.53) (56,72.53) (63,71.79) (70,70.7) (77,69.96) (84,69.23)};
\addlegendentry{2018}
\addplot[bends_color, mark=none] coordinates {(0,68.93) (1,67.7) (2,66.05) (3,65.64) (4,65.43) (5,64.81) (6,64.61) (7,63.99) (14,60.7) (21,58.64) (28,57) (35,55.56) (42,54.73) (49,51.85) (56,51.03) (63,50.62) (70,49.79) (77,49.18) (84,48.77)};
\addlegendentry{2021}
\addplot[mid_color, mark=none] coordinates {(0,52.1) (1,48.87) (2,46.13) (3,44.84) (4,44.19) (5,43.06) (6,42.58) (7,41.13) (14,37.58) (21,35.81) (28,34.52) (35,32.42) (42,30.65) (49,29.35) (56,27.74) (63,26.94) (70,26.29) (77,25) (84,24.35)};
\addlegendentry{2024}
\addplot[breaks_color, mark=none] coordinates {(0,46.38) (1,42.44) (2,40.79) (3,38.51) (4,37.89) (5,36.65) (6,34.58) (7,33.75) (14,31.47) (21,29.4) (28,26.5) (35,24.84) (42,23.6) (49,21.95) (56,20.91) (63,19.05) (70,17.18) (77,15.53) (84,14.91)};
\addlegendentry{2025}
\end{groupplot}
\end{tikzpicture}

%% file: tex/appendix_euvd.tex
\section{Cross-check against the European Vulnerability Database}
\label{app:euvd}

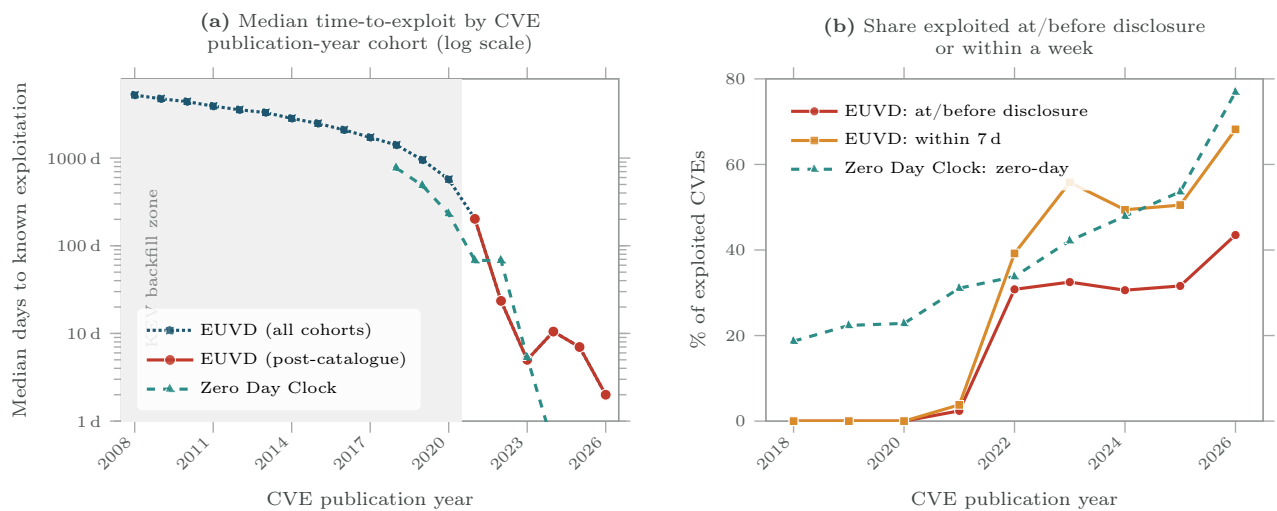
\begin{figure}[ht]
\centering
\resizebox{0.98\textwidth}{!}{\input{tex/fig_euvd.tex}}
\caption{\textbf{The EU catalogue reproduces the collapse under a different dating rule.} \textbf{(a)} Median time-to-known-exploitation by CVE publication-year cohort, log scale. The EUVD series (solid) falls from 202~days (2021) to 24~(2022) to 5~(2023) and single digits thereafter, the same order-of-magnitude collapse the Zero Day Clock series (dashed) shows. The shaded band marks cohorts published before the CISA KEV catalogue existed (Nov 2021), where the apparent decline is partly a backfill artefact and should not be over-read. \textbf{(b)} Share of exploited CVEs known to be exploited at or before disclosure, and within a week, by publication cohort; both rise from near zero to 30--68\%. Data: ENISA EUVD API, snapshot of 1~July 2026 ($n=1{,}632$).}
\label{fig:euvd}
\end{figure}

A single data source is a weak foundation for a structural claim, so we repeat the time-to-exploit analysis on the European Union Vulnerability Database (EUVD), operated by ENISA under the NIS2 Directive and CRA. We queried the EUVD API directly~\citep{euvd2026}, retrieved its full known-exploited-vulnerability dump (1{,}634 records), and fetched each record's publication date, obtaining 1{,}632 usable pairs. For each we measured time-to-exploit as the interval from CVE publication to the date it entered the exploited-vulnerability catalogue, binned by publication-year cohort. Figure~\ref{fig:euvd} overlays the result on the Zero Day Clock series from Appendix~\ref{app:zerodayclock}. (This is a separately maintained catalogue with its own dating rule but, as Appendix~\ref{app:euroclock} quantifies, its exploited feed overlaps almost entirely with the US CISA catalogue, so it is best read as the same data under a more conservative dating rule.)

\paragraph{Do the patterns match?} Yes, on the axis that matters. For cohorts published after the catalogue existed (the uncontaminated window), the EUVD median time-to-exploit falls from roughly 200~days to single digits within three years, and the share of flaws weaponised at or before disclosure rises from near zero to 30--43\%. Both movements reproduce the direction and approximate magnitude of the Zero Day Clock trend under a different dating rule.

\paragraph{Why the absolute numbers differ.} The EUVD medians sit \emph{above} the Zero Day Clock medians for recent cohorts, and the reason is methodological. EUVD dates a vulnerability from when it was \emph{added to the catalogue}, which lags the moment exploitation first occurred; the Zero Day Clock dates the first public exploit artefact. EUVD is thus a systematically conservative proxy: it over-states the window and still shows it closing to days. When the pessimistic and aggressive measures agree on the trend, the trend is robust.

\paragraph{A critical reading of the EUVD data.} The caveats again cut toward caution, not alarm.
\begin{itemize}
\item \textbf{Catalogue-inception backfill.} CISA KEV launched in November 2021 and EUVD in 2024, so older vulnerabilities could only be added later, mechanically inflating measured time-to-exploit for old cohorts. We shade that region in Figure~\ref{fig:euvd}(a) and rest no claim on it.
\item \textbf{Exploited-subset selection.} Like the Zero Day Clock, this corpus is conditioned on vulnerabilities \emph{known to be exploited}. It measures the tempo of the exploited subset, not the probability of exploitation.
\item \textbf{Small annual $n$ and coarse dating.} Per-cohort counts are in the low hundreds and dates are day-resolution, so single-year wiggles are noise; only the multi-year trend is interpretable.
\item \textbf{Provenance overlap is near-total.} The EUVD exploited feed is 99.8\% sourced from CISA KEV (Appendix~\ref{app:euroclock}, Finding~1): only 4 of 1{,}632 records are unique to ENISA's own feed. This is not an independent dataset that happens to overlap; it is nearly the same catalogue under a different dating rule. We therefore claim robustness to the dating rule, not replication by a disjoint source.
\end{itemize}
\noindent Read critically, the EUVD cross-check does not add precision; it adds \emph{robustness}. The finding that survives both dating conventions and the removal of every backfill-contaminated cohort is the one the argument needs: for vulnerabilities that are exploited, the window in which a point-in-time, patch-cycle conformity model could intervene has closed to days. Premise~P4 fails on EU-maintained data as surely as on the Zero Day Clock's.

%% file: tex/fig_euvd.tex
\begin{tikzpicture}
\begin{groupplot}[
  group style={group size=2 by 1, horizontal sep=1.9cm},
  width=8.0cm, height=6.0cm,
  tick label style={font=\scriptsize},
  label style={font=\footnotesize},
  title style={font=\footnotesize, align=center},
  legend style={font=\scriptsize, draw=none, fill=white, fill opacity=0.85, text opacity=1},
  every axis plot/.append style={line width=1.15pt, mark options={draw=white, line width=0.35pt}},
  x tick label style={/pgf/number format/1000 sep={}},
]
\nextgroupplot[
  title={\textbf{(a)} Median time-to-exploit by CVE\\publication-year cohort (log scale)},
  xlabel={CVE publication year}, ylabel={Median days to known exploitation},
  ymode=log, ymin=1, ymax=8000, xmin=2007.5, xmax=2026.5,
  xtick={2008,2011,2014,2017,2020,2023,2026}, xticklabel style={rotate=45,anchor=east},
  ytick={1,10,100,1000}, yticklabels={1\,d,10\,d,100\,d,1000\,d},
  legend pos=south west,
]
\fill[cai_dark!8] (axis cs:2007.5,1) rectangle (axis cs:2020.5,8000);
\node[cai_dark!70, font=\scriptsize, rotate=90, anchor=south] at (axis cs:2009.2,55) {KEV backfill zone};
\addplot[cra_color, mark=square*, mark size=1.5pt, densely dotted] coordinates {(2008,5229) (2009,4745.5) (2010,4402) (2011,3914) (2012,3555) (2013,3303) (2014,2837.0) (2015,2485) (2016,2097) (2017,1720.5) (2018,1409.0) (2019,951) (2020,572.0) (2021,202.5) (2022,23.5) (2023,5) (2024,10.5) (2025,7.0) (2026,2)};
\addlegendentry{EUVD (all cohorts)}
\addplot[breaks_color, mark=*, mark size=2pt] coordinates {(2021,202.5) (2022,23.5) (2023,5) (2024,10.5) (2025,7.0) (2026,2)};
\addlegendentry{EUVD (post-catalogue)}
\addplot[bends_color, mark=triangle*, mark size=2.2pt, dashed] coordinates {(2018,770.98) (2019,485.27) (2020,231.2) (2021,68.17) (2022,67.94) (2023,5.32) (2024,0.52)};
\addlegendentry{Zero Day Clock}
\nextgroupplot[
  title={\textbf{(b)} Share exploited at/before disclosure\\or within a week},
  xlabel={CVE publication year}, ylabel={\% of exploited CVEs}, ymin=0, ymax=80,
  xmin=2017.5, xmax=2026.5, xtick={2018,2020,2022,2024,2026},
  xticklabel style={rotate=45,anchor=east}, ytick={0,20,40,60,80},
  legend pos=north west,
]
\addplot[breaks_color, mark=*, mark size=1.7pt] coordinates {(2018,0.0) (2019,0.0) (2020,0.0) (2021,2.4) (2022,30.8) (2023,32.5) (2024,30.6) (2025,31.6) (2026,43.5)};
\addlegendentry{EUVD: at/before disclosure}
\addplot[mid_color, mark=square*, mark size=1.6pt] coordinates {(2018,0.0) (2019,0.0) (2020,0.0) (2021,3.8) (2022,39.2) (2023,55.8) (2024,49.4) (2025,50.5) (2026,68.2)};
\addlegendentry{EUVD: within 7\,d}
\addplot[bends_color, mark=triangle*, mark size=2pt, dashed] coordinates {(2018,18.68) (2019,22.37) (2020,22.84) (2021,31.07) (2022,33.78) (2023,42.18) (2024,47.9) (2025,53.62) (2026,76.92)};
\addlegendentry{Zero Day Clock: zero-day}
\end{groupplot}
\end{tikzpicture}

%% file: tex/appendix_euroclock.tex
\section{A European Zero Day Clock: deeper structure in the EUVD data}
\label{app:euroclock}

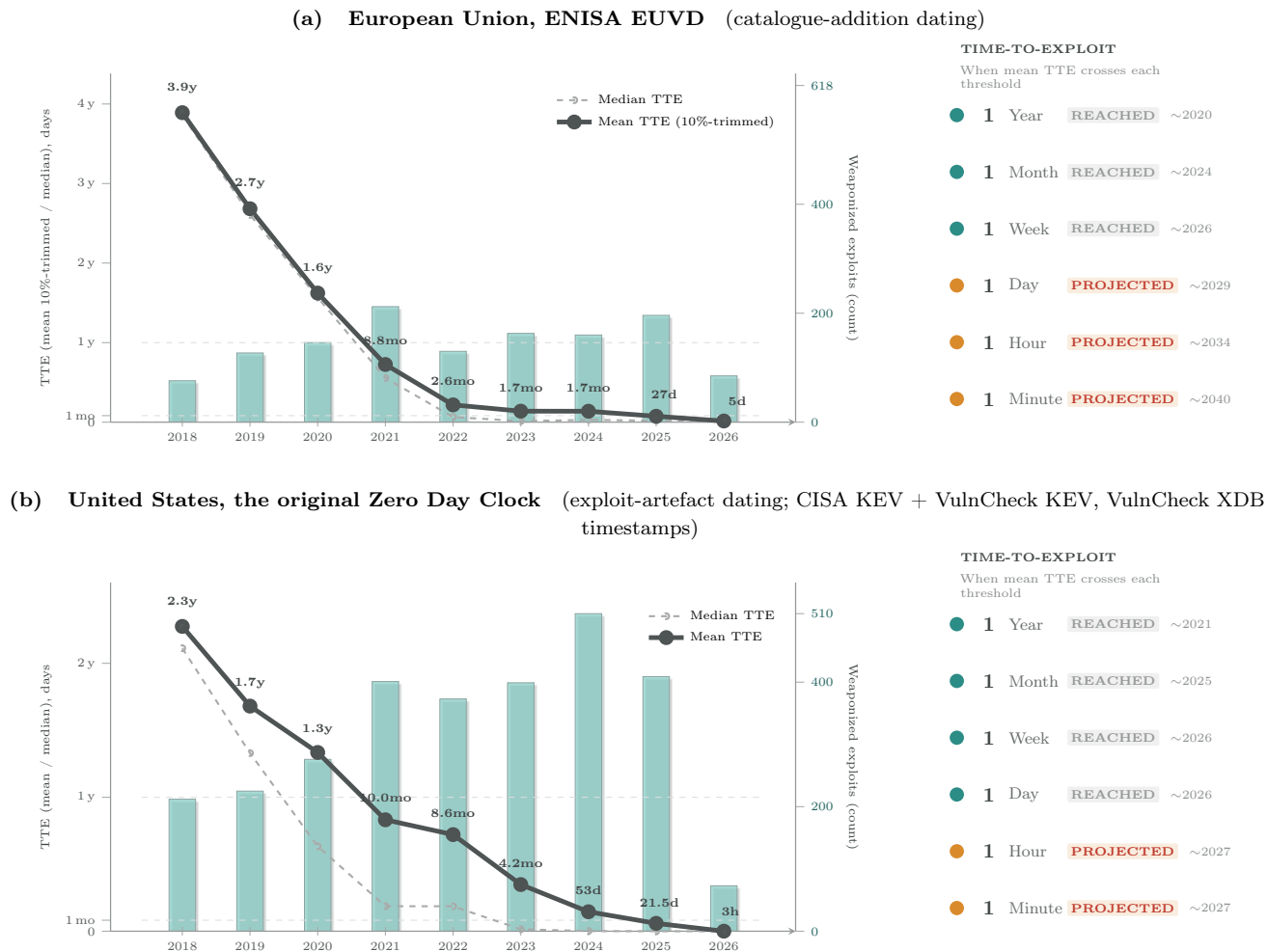
\begin{figure*}[t]
\centering
{\small\textbf{(a)\quad European Union, ENISA EUVD} \; (catalogue-addition dating)}\\[2pt]
\resizebox{0.96\textwidth}{!}{\input{tex/fig_euvd_zdclock.tex}}\\[12pt]
{\small\textbf{(b)\quad United States, the original Zero Day Clock} \; (exploit-artefact dating; CISA KEV + VulnCheck KEV, VulnCheck XDB timestamps)}\\[2pt]
\resizebox{0.96\textwidth}{!}{\input{tex/fig_us_zdclock.tex}}
\caption{\textbf{Two Zero Day Clocks, one collapse.} The same visualisation applied to two exploited-vulnerability datasets: bold line = mean time-to-exploit by CVE publication-year cohort, dashed line = median, bars (right axis) = weaponised-exploit count, and the sidebar marks when the mean-TTE series crosses each threshold. \textbf{(a)} ENISA's European Vulnerability Database ($n=1{,}632$), dated at catalogue addition: the mean falls from 3.9~years (2018) to 5~days in the 2026 cohort, so the one-year, one-month, and one-week thresholds are already crossed, with only the one-day line and below still projected. \textbf{(b)} The original Zero Day Clock~\citep{zerodayclock2026}, dated at the first public exploit artefact: the mean falls from 2.3~years (2018) to about 3~hours in the partial 2026 cohort, with the one-year (2021), one-month (2025), one-week (2026), and one-day (2026) thresholds already reached and one-hour/one-minute projected. The per-cohort value labels and sidebar dates reproduce the live zerodayclock.com dashboard as of the snapshot date. Panel~(b)'s milestones run ahead of~(a)'s because exploit-artefact dating registers exploitation earlier than a catalogue does, so the two panels bound the same collapse from its fast and slow sides. The two feeds are not independent (99.8\% of the EUVD records also appear in CISA KEV, Finding~1), so this is one collapse measured under two dating rules, not two separate measurements. Data snapshots: 1~July 2026.}
\label{fig:clocks}
\end{figure*}

\begin{figure}[ht]
\centering
\resizebox{0.98\textwidth}{!}{\input{tex/fig_euroclock.tex}}
\caption{\textbf{The failure of triage instruments.} All panels use ENISA EUVD data only ($n=1{,}632$). \textbf{(a)} Median time-to-exploit by publication cohort (log scale) with an exponential-decay fit ($R^2=0.90$), whose central estimate crosses a one-day median in 2026 and a one-hour median in 2030. \textbf{(b)} Median time-to-exploit by EPSS predicted-exploit-probability band (post-2021 cohorts): the relationship is \emph{inverted}: vulnerabilities the predictor rates most likely to be exploited are exploited \emph{more slowly} (Spearman $\rho=+0.15$, $p<10^{-5}$), and 42\% of vulnerabilities exploited within a day scored below 10\%. \textbf{(c)} Median time-to-exploit by CVSS severity band across three eras: near 550~days for every band in 2019--2021, collapsing to a narrow 3--11-day range by 2024--2026, so severity no longer orders exploitation tempo. Data: ENISA EUVD API, snapshot 1~July 2026.}
\label{fig:euroclock}
\end{figure}
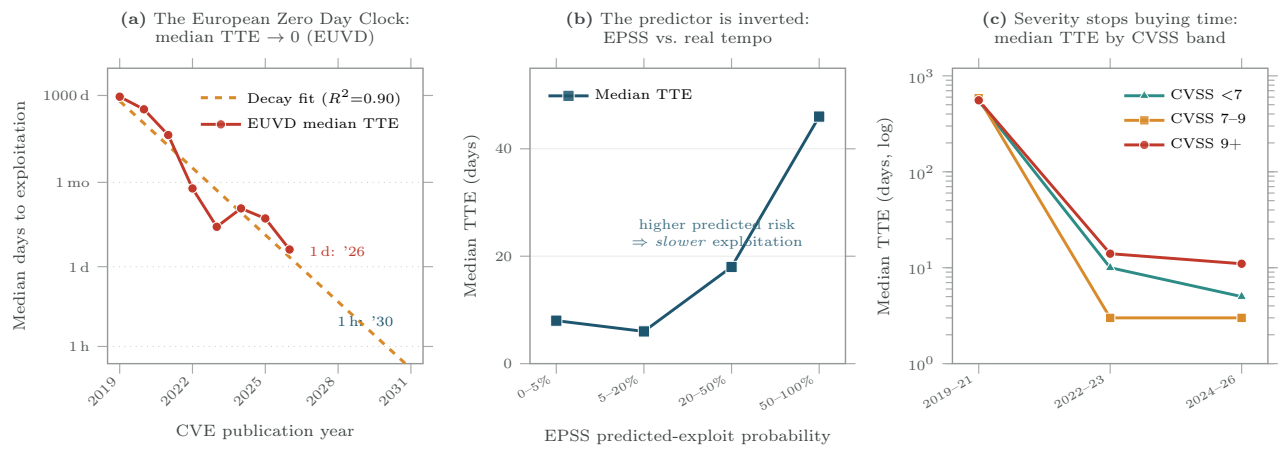

Appendix~\ref{app:euvd} showed that the EUVD reproduces the time-to-exploit collapse. Here we go further. Figure~\ref{fig:clocks} redraws the EUVD record in the Zero Day Clock's own visual idiom and places the original US clock directly beneath it for comparison; Figure~\ref{fig:euroclock} then asks whether the risk-scoring instruments a manufacturer would use to triage vulnerabilities still work.

\paragraph{Finding 1: two clocks, one collapse, measured under two dating rules, not from two independent datasets.} Provenance first, because it bounds the claim. Of the 1{,}632 exploited EUVD records here, 1{,}628 (99.8\%) also appear in the US CISA Known Exploited Vulnerabilities catalogue; only four are unique to ENISA's own feed. The EUVD exploited set is therefore very nearly the US catalogue re-served through an EU API, not an independently assembled European dataset, and our earlier description of it as a ``wholly separate corpus'' (Appendix~\ref{app:euvd}) overstated the case. What genuinely differs between the two panels of Figure~\ref{fig:clocks} is the \emph{dating rule} (EUVD dates catalogue addition, the Zero Day Clock dates the first public exploit) and the \emph{source breadth}. Those differences make the EUVD clock the slower of the two (one-day median around 2029 against 2028), so the pair brackets the collapse from a conservative and an aggressive side. The conclusion that survives is simple: under both dating rules, mean time-to-exploit has fallen from years to days and has already crossed the one-month line. A truly independent European signal will exist only once ENISA's feed accumulates enough non-CISA entries to stand alone; it cannot yet, and we do not claim otherwise.

\paragraph{Finding 2: the risk-scoring instruments the \cra{} relies on fail where they matter.} The \cra{}'s risk-based vulnerability handling (Annex~I, Part~II) assumes a manufacturer can prioritise, that is, find the vulnerabilities most likely to be exploited and fix them first. That needs a working predictor. On EUVD data, the two the industry uses both fail (Figure~\ref{fig:euroclock}b,c). EPSS, an exploitation-probability score, is not merely uninformative but \emph{inverted}: the vulnerabilities it rates highest-risk are exploited later, and nearly half of the fastest-exploited flaws scored low. CVSS severity fares no better: by 2024--2026 every severity band is exploited within a 3--11-day window, so ``patch the critical ones first'' buys almost no time.

\paragraph{Why this deepens the \emph{breaks} argument.} A natural defence of the risk-based model is that manufacturers need not fix \emph{everything}: abundance is tolerable if they can triage. That assumes triage is possible. Finding~2 says it is not: at machine tempo the triage instruments have decoupled from the outcome they predict, so a risk-based process cannot prioritise its way out of the tempo collapse. The bends remedy (re-read the obligation as demonstrable prioritisation) and the breaks problem (the environment changed) are therefore linked: the collapse of premise~P3 also removes the signal the P1 remedy would need.

\paragraph{Caveats.} These are cross-sectional associations on the exploited subset, and the limits from Appendices~\ref{app:zerodayclock} and~\ref{app:euvd} carry over. In particular, EPSS and CVSS scores are read at snapshot time, so a score revised upward \emph{after} exploitation would make the predictor look better than it was in advance, meaning the inversion is, if anything, understated. We therefore read Finding~2 as a robust \emph{direction} (the instruments do not order tempo; EPSS is at best uninformative for the fastest attacks) and treat the exact coefficients as indicative. The modest conclusion suffices: at this tempo a manufacturer cannot know in advance which vulnerabilities will be exploited first, so a prioritise-then-patch conformity process runs without the signal it presupposes.

%% file: tex/fig_euvd_zdclock.tex
\begin{tikzpicture}
\begin{axis}[
  name=main, width=13.4cm, height=7.6cm,
  axis y line*=right, axis x line=none,
  ymin=0, ymax=640, xmin=2017.4, xmax=2026.6,
  ylabel={Weaponized exploits (count)},
  ylabel style={font=\scriptsize\color{bends_color!80!black}}, y label style={rotate=180},
  ytick={0,200,400,618}, yticklabel style={font=\scriptsize,color=bends_color!70!black},
  enlarge x limits=0.05,
]
\addplot[ybar, bar width=13pt, draw=graph_navy!55, line width=0.4pt, fill=bends_light, fill opacity=0.68,
  drop shadow={shadow xshift=1.3pt, shadow yshift=-1.3pt, fill=cai_dark, opacity=0.22}]
  coordinates {(2018,76) (2019,127) (2020,146) (2021,212) (2022,130) (2023,163) (2024,160) (2025,196) (2026,85)};
\end{axis}
\begin{axis}[
  name=lines, width=13.4cm, height=7.6cm, at=(main.south west), anchor=south west,
  axis y line*=left, axis x line=bottom,
  ymin=0, ymax=1600, xmin=2017.4, xmax=2026.6,
  xtick={2018,2019,2020,2021,2022,2023,2024,2025,2026},
  xticklabel style={font=\scriptsize}, x tick label style={/pgf/number format/1000 sep={}},
  ylabel={TTE (mean 10\%-trimmed / median), days},
  ylabel style={font=\scriptsize}, ytick={0,30,365,730,1095,1460},
  yticklabels={0,1\,mo,1\,y,2\,y,3\,y,4\,y}, yticklabel style={font=\scriptsize},
  enlarge x limits=0.05, clip=false,
  legend style={at={(0.985,0.97)},anchor=north east,font=\scriptsize,draw=none,
    fill=white,fill opacity=0.85,text opacity=1,row sep=1pt},
]
\draw[cai_dark!18,dashed] (axis cs:2017.4,30)--(axis cs:2026.6,30);
\draw[cai_dark!18,dashed] (axis cs:2017.4,365)--(axis cs:2026.6,365);
\addplot[cai_dark!45, line width=1pt, dashed, mark=o, mark size=1.4pt,
  mark options={fill=white}] coordinates {(2018,1409.0) (2019,951.0) (2020,572.0) (2021,202.5) (2022,23.5) (2023,5.0) (2024,10.5) (2025,7.0) (2026,2.0)};
\addlegendentry{Median TTE}
\addplot[cai_dark!92, line width=2.3pt, mark=*, mark size=2.5pt] coordinates {(2018,1420.2) (2019,979.3) (2020,592.4) (2021,264.7) (2022,79.4) (2023,50.5) (2024,50.2) (2025,27.1) (2026,5.4)};
\addlegendentry{Mean TTE (10\%-trimmed)}
\node[font=\scriptsize\bfseries,cai_dark,anchor=south] at (axis cs:2018.0,1480.225806451613) {3.9y};
\node[font=\scriptsize\bfseries,cai_dark,anchor=south] at (axis cs:2019.0,1039.3203883495146) {2.7y};
\node[font=\scriptsize\bfseries,cai_dark,anchor=south] at (axis cs:2020.0,652.4491525423729) {1.6y};
\node[font=\scriptsize\bfseries,cai_dark,anchor=south] at (axis cs:2021.0,326.74117647058824) {8.8mo};
\node[font=\scriptsize\bfseries,cai_dark,anchor=south] at (axis cs:2022.0,139.3653846153846) {2.6mo};
\node[font=\scriptsize\bfseries,cai_dark,anchor=south] at (axis cs:2023.0,108.45038167938932) {1.7mo};
\node[font=\scriptsize\bfseries,cai_dark,anchor=south] at (axis cs:2024.0,108.1953125) {1.7mo};
\node[font=\scriptsize\bfseries,cai_dark,anchor=south] at (axis cs:2025.12,79.1139240506329) {27d};
\node[font=\scriptsize\bfseries,cai_dark,anchor=south] at (axis cs:2026.2,47.36231884057971) {5d};

\end{axis}
\begin{scope}[shift={(14.6,5.3)}]
\node[anchor=west,font=\scriptsize\bfseries,cai_dark] at (-0.05,1.15) {TIME-TO-EXPLOIT};
\node[anchor=west,font=\scriptsize,cai_dark!62,text width=3.7cm,align=left] at (-0.05,0.66)
  {When mean TTE crosses each threshold};
\node[circle,fill=bends_color,minimum size=7pt,inner sep=0] (d0) at (0,0.0) {};
\node[anchor=west,font=\large\bfseries,cai_dark] at (0.34,0.0) {1};
\node[anchor=west,font=\small,cai_dark!80] at (0.78,0.0) {Year};
\node[anchor=west,font=\scriptsize\bfseries,text=cai_dark!55,fill=cai_dark!8,inner sep=1.8pt,rounded corners=1pt] (t0) at (1.9,0.0) {REACHED};
\node[anchor=west,font=\scriptsize,cai_dark!60] at (t0.east) {\;$\sim$2020};
\node[circle,fill=bends_color,minimum size=7pt,inner sep=0] (d1) at (0,-0.98) {};
\node[anchor=west,font=\large\bfseries,cai_dark] at (0.34,-0.98) {1};
\node[anchor=west,font=\small,cai_dark!80] at (0.78,-0.98) {Month};
\node[anchor=west,font=\scriptsize\bfseries,text=cai_dark!55,fill=cai_dark!8,inner sep=1.8pt,rounded corners=1pt] (t1) at (1.9,-0.98) {REACHED};
\node[anchor=west,font=\scriptsize,cai_dark!60] at (t1.east) {\;$\sim$2024};
\node[circle,fill=bends_color,minimum size=7pt,inner sep=0] (d2) at (0,-1.96) {};
\node[anchor=west,font=\large\bfseries,cai_dark] at (0.34,-1.96) {1};
\node[anchor=west,font=\small,cai_dark!80] at (0.78,-1.96) {Week};
\node[anchor=west,font=\scriptsize\bfseries,text=cai_dark!55,fill=cai_dark!8,inner sep=1.8pt,rounded corners=1pt] (t2) at (1.9,-1.96) {REACHED};
\node[anchor=west,font=\scriptsize,cai_dark!60] at (t2.east) {\;$\sim$2026};
\node[circle,fill=mid_color,minimum size=7pt,inner sep=0] (d3) at (0,-2.94) {};
\node[anchor=west,font=\large\bfseries,cai_dark] at (0.34,-2.94) {1};
\node[anchor=west,font=\small,cai_dark!80] at (0.78,-2.94) {Day};
\node[anchor=west,font=\scriptsize\bfseries,text=breaks_color,fill=mid_color!15,inner sep=1.8pt,rounded corners=1pt] (t3) at (1.9,-2.94) {PROJECTED};
\node[anchor=west,font=\scriptsize,cai_dark!60] at (t3.east) {\;$\sim$2029};
\node[circle,fill=mid_color,minimum size=7pt,inner sep=0] (d4) at (0,-3.92) {};
\node[anchor=west,font=\large\bfseries,cai_dark] at (0.34,-3.92) {1};
\node[anchor=west,font=\small,cai_dark!80] at (0.78,-3.92) {Hour};
\node[anchor=west,font=\scriptsize\bfseries,text=breaks_color,fill=mid_color!15,inner sep=1.8pt,rounded corners=1pt] (t4) at (1.9,-3.92) {PROJECTED};
\node[anchor=west,font=\scriptsize,cai_dark!60] at (t4.east) {\;$\sim$2034};
\node[circle,fill=mid_color,minimum size=7pt,inner sep=0] (d5) at (0,-4.9) {};
\node[anchor=west,font=\large\bfseries,cai_dark] at (0.34,-4.9) {1};
\node[anchor=west,font=\small,cai_dark!80] at (0.78,-4.9) {Minute};
\node[anchor=west,font=\scriptsize\bfseries,text=breaks_color,fill=mid_color!15,inner sep=1.8pt,rounded corners=1pt] (t5) at (1.9,-4.9) {PROJECTED};
\node[anchor=west,font=\scriptsize,cai_dark!60] at (t5.east) {\;$\sim$2040};
\end{scope}
\end{tikzpicture}

%% file: tex/fig_us_zdclock.tex
\begin{tikzpicture}
\begin{axis}[
  name=main, width=13.4cm, height=7.6cm, axis y line*=right, axis x line=none,
  ymin=0, ymax=560, xmin=2017.4, xmax=2026.6,
  ylabel={Weaponized exploits (count)},
  ylabel style={font=\scriptsize\color{bends_color!80!black}}, y label style={rotate=180},
  ytick={0,200,400,510}, yticklabel style={font=\scriptsize,color=bends_color!70!black},
  enlarge x limits=0.05,
]
\addplot[ybar, bar width=13pt, draw=graph_navy!55, line width=0.4pt, fill=bends_light, fill opacity=0.68,
  drop shadow={shadow xshift=1.3pt, shadow yshift=-1.3pt, fill=cai_dark, opacity=0.22}]
  coordinates {(2018,212) (2019,225) (2020,276) (2021,401) (2022,373) (2023,399) (2024,510) (2025,409) (2026,73)};
\end{axis}
\begin{axis}[
  name=lines, width=13.4cm, height=7.6cm, at=(main.south west), anchor=south west,
  axis y line*=left, axis x line=bottom, ymin=0, ymax=950, xmin=2017.4, xmax=2026.6,
  xtick={2018,2019,2020,2021,2022,2023,2024,2025,2026},
  xticklabel style={font=\scriptsize}, x tick label style={/pgf/number format/1000 sep={}},
  ylabel={TTE (mean / median), days}, ylabel style={font=\scriptsize},
  ytick={0,30,365,730}, yticklabels={0,1\,mo,1\,y,2\,y}, yticklabel style={font=\scriptsize},
  enlarge x limits=0.05, clip=false,
  legend style={at={(0.985,0.95)},anchor=north east,font=\scriptsize,draw=none,
    fill=white,fill opacity=0.85,text opacity=1,row sep=1pt},
]
\draw[cai_dark!18,dashed] (axis cs:2017.4,30)--(axis cs:2026.6,30);
\draw[cai_dark!18,dashed] (axis cs:2017.4,365)--(axis cs:2026.6,365);
\addplot[cai_dark!45, line width=1pt, dashed, mark=o, mark size=1.4pt,
  mark options={fill=white}] coordinates {(2018,770.98) (2019,485.27) (2020,231.2) (2021,68.17) (2022,67.94) (2023,5.32) (2024,0.52) (2025,0) (2026,0)};
\addlegendentry{Median TTE}
\addplot[cai_dark!92, line width=2.3pt, mark=*, mark size=2.5pt] coordinates {(2018,830.42) (2019,613.30) (2020,487.02) (2021,303.59) (2022,263.21) (2023,127.15) (2024,53.08) (2025,21.47) (2026,0.12)};
\addlegendentry{Mean TTE}
\node[font=\scriptsize\bfseries,cai_dark,anchor=south] at (axis cs:2018,864.42) {2.3y};
\node[font=\scriptsize\bfseries,cai_dark,anchor=south] at (axis cs:2019,645.3) {1.7y};
\node[font=\scriptsize\bfseries,cai_dark,anchor=south] at (axis cs:2020,519.02) {1.3y};
\node[font=\scriptsize\bfseries,cai_dark,anchor=south] at (axis cs:2021,335.59) {10.0mo};
\node[font=\scriptsize\bfseries,cai_dark,anchor=south] at (axis cs:2022,293.21) {8.6mo};
\node[font=\scriptsize\bfseries,cai_dark,anchor=south] at (axis cs:2023,157.15) {4.2mo};
\node[font=\scriptsize\bfseries,cai_dark,anchor=south] at (axis cs:2024,83.08) {53d};
\node[font=\scriptsize\bfseries,cai_dark,anchor=south] at (axis cs:2025.05,49.47) {21.5d};
\node[font=\scriptsize\bfseries,cai_dark,anchor=south] at (axis cs:2026.1,26) {3h};

\end{axis}
\begin{scope}[shift={(14.6,5.3)}]
\node[anchor=west,font=\scriptsize\bfseries,cai_dark] at (-0.05,1.15) {TIME-TO-EXPLOIT};
\node[anchor=west,font=\scriptsize,cai_dark!62,text width=3.7cm,align=left] at (-0.05,0.66)
  {When mean TTE crosses each threshold};
\node[circle,fill=bends_color,minimum size=7pt,inner sep=0] (d0) at (0,0.0) {};
\node[anchor=west,font=\large\bfseries,cai_dark] at (0.34,0.0) {1};
\node[anchor=west,font=\small,cai_dark!80] at (0.78,0.0) {Year};
\node[anchor=west,font=\scriptsize\bfseries,text=cai_dark!55,fill=cai_dark!8,inner sep=1.8pt,rounded corners=1pt] (t0) at (1.9,0.0) {REACHED};
\node[anchor=west,font=\scriptsize,cai_dark!60] at (t0.east) {\;$\sim$2021};
\node[circle,fill=bends_color,minimum size=7pt,inner sep=0] (d1) at (0,-0.98) {};
\node[anchor=west,font=\large\bfseries,cai_dark] at (0.34,-0.98) {1};
\node[anchor=west,font=\small,cai_dark!80] at (0.78,-0.98) {Month};
\node[anchor=west,font=\scriptsize\bfseries,text=cai_dark!55,fill=cai_dark!8,inner sep=1.8pt,rounded corners=1pt] (t1) at (1.9,-0.98) {REACHED};
\node[anchor=west,font=\scriptsize,cai_dark!60] at (t1.east) {\;$\sim$2025};
\node[circle,fill=bends_color,minimum size=7pt,inner sep=0] (d2) at (0,-1.96) {};
\node[anchor=west,font=\large\bfseries,cai_dark] at (0.34,-1.96) {1};
\node[anchor=west,font=\small,cai_dark!80] at (0.78,-1.96) {Week};
\node[anchor=west,font=\scriptsize\bfseries,text=cai_dark!55,fill=cai_dark!8,inner sep=1.8pt,rounded corners=1pt] (t2) at (1.9,-1.96) {REACHED};
\node[anchor=west,font=\scriptsize,cai_dark!60] at (t2.east) {\;$\sim$2026};
\node[circle,fill=bends_color,minimum size=7pt,inner sep=0] (d3) at (0,-2.94) {};
\node[anchor=west,font=\large\bfseries,cai_dark] at (0.34,-2.94) {1};
\node[anchor=west,font=\small,cai_dark!80] at (0.78,-2.94) {Day};
\node[anchor=west,font=\scriptsize\bfseries,text=cai_dark!55,fill=cai_dark!8,inner sep=1.8pt,rounded corners=1pt] (t3) at (1.9,-2.94) {REACHED};
\node[anchor=west,font=\scriptsize,cai_dark!60] at (t3.east) {\;$\sim$2026};
\node[circle,fill=mid_color,minimum size=7pt,inner sep=0] (d4) at (0,-3.92) {};
\node[anchor=west,font=\large\bfseries,cai_dark] at (0.34,-3.92) {1};
\node[anchor=west,font=\small,cai_dark!80] at (0.78,-3.92) {Hour};
\node[anchor=west,font=\scriptsize\bfseries,text=breaks_color,fill=mid_color!15,inner sep=1.8pt,rounded corners=1pt] (t4) at (1.9,-3.92) {PROJECTED};
\node[anchor=west,font=\scriptsize,cai_dark!60] at (t4.east) {\;$\sim$2027};
\node[circle,fill=mid_color,minimum size=7pt,inner sep=0] (d5) at (0,-4.9) {};
\node[anchor=west,font=\large\bfseries,cai_dark] at (0.34,-4.9) {1};
\node[anchor=west,font=\small,cai_dark!80] at (0.78,-4.9) {Minute};
\node[anchor=west,font=\scriptsize\bfseries,text=breaks_color,fill=mid_color!15,inner sep=1.8pt,rounded corners=1pt] (t5) at (1.9,-4.9) {PROJECTED};
\node[anchor=west,font=\scriptsize,cai_dark!60] at (t5.east) {\;$\sim$2027};
\end{scope}
\end{tikzpicture}

%% file: tex/fig_euroclock.tex
\begin{tikzpicture}
\begin{groupplot}[
  group style={group size=3 by 1, horizontal sep=1.6cm},
  width=6.3cm, height=6.0cm,
  tick label style={font=\scriptsize}, label style={font=\footnotesize},
  title style={font=\footnotesize, align=center},
  legend style={font=\scriptsize, draw=none, fill=white, fill opacity=0.85, text opacity=1},
  every axis plot/.append style={line width=1.25pt, mark options={draw=white, line width=0.35pt}},
  x tick label style={/pgf/number format/1000 sep={}},
]
\nextgroupplot[
  title={\textbf{(a)} The European Zero Day Clock:\\median TTE $\to 0$ (EUVD)},
  xlabel={CVE publication year}, ylabel={Median days to exploitation},
  ymode=log, ymin=0.02, ymax=3000, xmin=2018.5, xmax=2031.5,
  xtick={2019,2022,2025,2028,2031}, xticklabel style={rotate=45,anchor=east},
  ytick={0.04,1,30,1000}, yticklabels={1\,h,1\,d,1\,mo,1000\,d},
  legend pos=north east,
]
\draw[cai_dark!30,dotted] (axis cs:2018.5,30)--(axis cs:2031.5,30);
\draw[cai_dark!30,dotted] (axis cs:2018.5,1)--(axis cs:2031.5,1);
\draw[cai_dark!30,dotted] (axis cs:2018.5,0.04)--(axis cs:2031.5,0.04);
\addplot[mid_color,dashed,line width=1.3pt,domain=2019:2031,samples=80]{1966.232*exp(-0.89991*(x-2018))};
\addlegendentry{Decay fit ($R^2{=}0.90$)}
\addplot[breaks_color,mark=*,mark size=2pt] coordinates {(2019,951.0) (2020,572.0) (2021,202.5) (2022,23.5) (2023,5.0) (2024,10.5) (2025,7.0) (2026,2.0)};
\addlegendentry{EUVD median TTE}
\node[breaks_color,font=\scriptsize,anchor=west] at (axis cs:2026.5,1.9){1\,d: '26};
\node[cra_color,font=\scriptsize,anchor=east] at (axis cs:2030.6,0.11){1\,h: '30};
\nextgroupplot[
  title={\textbf{(b)} The predictor is inverted:\\EPSS vs.\ real tempo},
  xlabel={EPSS predicted-exploit probability}, ylabel={Median TTE (days)},
  ymin=0,ymax=55, xtick={0,1,2,3}, xticklabels={0--5\%,5--20\%,20--50\%,50--100\%},
  xticklabel style={rotate=30,anchor=east,font=\tiny}, ymajorgrids,
  legend pos=north west,
]
\addplot[cra_color,mark=square*,mark size=2.4pt] coordinates {(0,8) (1,6) (2,18) (3,46)};
\addlegendentry{Median TTE}
\node[cra_color!85,font=\scriptsize,align=center,anchor=south east] at (axis cs:2.9,20)
  {higher predicted risk\\$\Rightarrow$ \emph{slower} exploitation};
\nextgroupplot[
  title={\textbf{(c)} Severity stops buying time:\\median TTE by CVSS band},
  xlabel={}, ylabel={Median TTE (days, log)}, ymode=log, ymin=1, ymax=1200,
  xtick={0,1,2}, xticklabels={2019--21,2022--23,2024--26},
  xticklabel style={rotate=30,anchor=east,font=\tiny}, ytick={1,10,100,1000},
  legend pos=north east,
]
\addplot[bends_color,mark=triangle*,mark size=2.4pt] coordinates {(0,548) (1,10) (2,5)};
\addlegendentry{CVSS $<$7}
\addplot[mid_color,mark=square*,mark size=2pt] coordinates {(0,584) (1,3) (2,3)};
\addlegendentry{CVSS 7--9}
\addplot[breaks_color,mark=*,mark size=2pt] coordinates {(0,556) (1,14) (2,11)};
\addlegendentry{CVSS 9+}
\end{groupplot}
\end{tikzpicture}